\documentclass[a4paper,12pt]{article}
\usepackage{graphicx}
\usepackage{hyperref}
\usepackage{latexsym}
\usepackage{color}
\usepackage{amsmath}
\usepackage{geometry}
\def\Tr{{\rm Tr}}
\def\msbar{{\overline{\rm MS}}}

\title{Non-perturbative renormalization of overlap quark bilinears on 2+1-flavor domain wall fermion configurations}
\author{Zhaofeng Liu$^1$, Ying Chen$^1$, Shao-Jing Dong$^2$, Michael Glatzmaier$^2$,\\
Ming Gong$^2$, Anyi Li$^3$, Keh-Fei Liu$^2$, Yi-Bo Yang$^1$ and Jian-Bo Zhang$^4$\\
($\chi$QCD Collaboration)}

\date{}
\begin{document}
\maketitle

\begin{center}
$^1$Institute of High Energy Physics and Theoretical Physics Center for Science Facilities, Chinese Academy of Sciences, Beijing 100049, China\\
$^2$Department of Physics and Astronomy, University of Kentucky, Lexington, KY 40506\\
$^3$Institute for Nuclear Theory, University of Washington, Seattle, WA 98195\\
$^4$Department of Physics, Zhejiang University, Hangzhou 311027, China
\end{center}

\begin{abstract}
We present renormalization constants of overlap quark bilinear operators on 2+1-flavor domain wall fermion configurations. 
This setup is being used by the $\chi$QCD collaboration in calculations of physical quantities such as strangeness in the nucleon 
and the strange and charm quark masses. 
The scale independent renormalization
constant for the axial vector current is computed using Ward Identity. The renormalization constants for scalar, pseudoscalar and vector current are 
calculated in the RI-MOM scheme. Results in the $\msbar$ scheme are also given. The step scaling function of quark masses in the RI-MOM scheme
is computed as well. The analysis uses in total six different ensembles of three sea quarks each on two lattices with sizes $24^3\times64$
and $32^3\times64$ at spacings $a = (1.73 {\rm~GeV})^{-1}$ and $ (2.28 {\rm~GeV})^{-1}$, respectively.
\end{abstract}

\newpage
\section{Introduction}
The overlap valence quark on $2+1$ flavor domain wall fermion (DWF) configurations
      has been used to calculate the strangeness and charmness in the nucleon~\cite{Gong:2013vja} with high precision. 
      Due to the high degree of chiral symmetry of these fermions, the calculation of the strangeness content is free of the problem
      of large mixing with the $\bar{u}u$ and $\bar{d}d$ matrix elements due to the additive renormalization of the quark mass
      that plagues the Wilson fermions. In addition to having small $\mathcal{O}(a^2)$ discretization errors~\cite{dll00,Draper:2005mh}, the overlap 
      fermion that we use for the valence quarks in the nucleon can also be used for the light and charm quarks
      with small $\mathcal{O}(m^2a^2)$ error~\cite{ld05,Li:2010pw}. This allows us to calculate the charmonium and charm-light
mesons in addition to strangeness and charmness contents. 
The inversion of overlap fermions can be speeded up 
by using HYP smearing~\cite{Hasenfratz:2001hp} and deflation with low eigenmodes~\cite{Li:2010pw}. 
The $\chi$QCD collaboration is determining charm and strange quark masses~\cite{Yang:2014taa}
and other physical quantities with the setup of overlap valence on DWF sea.
The renormalization constants of quark bilinear operators needed to match lattice results to those in the continuum $\msbar$ scheme are presented in this paper.

Non-perturbative renormalization is important in current lattice calculations aiming at percent level accuracy. As we know, the convergence of lattice perturbative calculations is often not satisfying and
lattice perturbation series rarely extend beyond the one-loop level.

We use the RI-MOM scheme~\cite{Martinelli:1994ty} to calculate renormalization constants for flavor non-singlet scalar, pseudoscalar, vector and axial vector operators
$\mathcal{O}=\bar\psi\Gamma\psi'$, where $\Gamma=I, \gamma_5, \gamma_\mu, \gamma_\mu\gamma_5$ respectively (we will use $S,P,V,A$ to denote the four operators throughout
this paper).
The results are converted to the $\msbar$ scheme using ratios from continuum perturbative calculations.
Following Refs.~\cite{Arthur:2010ht,Arthur:2013bqa}, we also calculate the step scaling function in the RI-MOM scheme for quark masses.
In this way, the $\mathcal{O}(a^2)$ discretization errors are removed differently.

We have calculated the renormalization constants at two lattice spacings with $a^{-1}=1.73(3)$ GeV and $2.28(3)$ GeV. At each lattice spacing, 
there are three light sea quark masses. At each light sea quark mass, we use eight valence quark masses. The final results are obtained in the chiral limit of
both the sea and valence quark masses, which confirm $Z_S=Z_P$ and $Z_V=Z_A$ for overlap fermions. The main results of this work are given in 
Tabs.~\ref{tab:Za_WI_24_32},~\ref{tab:zs},~\ref{tab:zs_error},~\ref{tab:zs_ri},~\ref{tab:zp} and~\ref{tab:zp_error}.

We consider the systematic errors carefully. A main source of systematic errors for $Z_S$ comes from the truncation of the 
perturbative ratio from the RI-MOM scheme to the $\msbar$ scheme.
We obtain $Z_S^\msbar(2$ GeV)=1.127(9)(19) on the coarse lattice and 1.056(6)(24) on the fine lattice, where the first uncertainty
is statistical and the second systematic.

This paper is organized as follows. In Sec.~\ref{sec:ri_and_overlap}, we briefly review the RI-MOM scheme and the overlap formalism. 
The numerical results in the RI-MOM and $\msbar$ schemes, the analysis of systematic errors and the calculation of the step scaling function
are given in Sec.~\ref{sec:num_results}. Then we summarize and conclude with some general remarks in Sec.~\ref{sec:summary}.

\section{Methodology}\label{sec:ri_and_overlap}
The non-perturbative calculation of renormalization constants in the RI-MOM scheme~\cite{Martinelli:1994ty} 
is based on imposing renormalization conditions on amputated Green functions of
the relevant operators in the momentum space. The Green functions needed to be computed include the quark propagator
\begin{equation}
S(p)=\sum_x e^{-ipx}\langle\psi(x)\bar\psi(0)\rangle,
\end{equation}
the forward Green function
\begin{equation}
G_{\mathcal{O}}(p)=\sum_{x,y}e^{-ip\cdot
(x-y)}\langle\psi(x)\mathcal{O}(0)\bar{\psi}(y)\rangle,
\end{equation}
and the vertex function
\begin{equation}
\Lambda_{\mathcal{O}}(p)=S^{-1}(p)G_{\mathcal{O}}(p)S^{-1}(p).
\end{equation}
The renormalization condition requires that the renormalized vertex function at a given scale $p^2=\mu^2$ coincides with its tree-level value. 
That is to say
\begin{equation}
Z_q^{-1}Z_{\mathcal{O}}\frac{1}{12}
\Tr\left[\Lambda_{\mathcal{O}}(p)\Lambda_{\mathcal{O}}^{tree}(p)^{-1}\right]_{p^2=\mu^2}=1,
\label{eq:ri_condition}
\end{equation}
where $Z_q$ is the quark field renormalization constant with $\psi_R=Z_q^{1/2}\psi$ (the subscript ``R" means after renormalization)
and $Z_{\mathcal{O}}$ the renormalization constant for operator $\mathcal{O}$ with
$\mathcal{O}_R=Z_{\mathcal{O}}\mathcal{O}$. Eq.(\ref{eq:ri_condition}) is defined in the quark massless limit so that RI-MOM is a mass
independent renormalization scheme. In practice, we do calculations at finite quark masses and then extrapolate to the chiral limit.
For convenience, a projected vertex function is defined by
\begin{equation}
\Gamma_{\mathcal{O}}(p)\equiv\frac{1}{12}\Tr[\Lambda_{\mathcal{O}}(p)\Lambda_{\mathcal{O}}^{tree}(p)^{-1}].
\label{eq:GammaS}
\end{equation}

In the RI scheme, $Z_q^{RI}$ can be determined by~\cite{Martinelli:1994ty}
\begin{equation}
Z_q^{RI}(\mu)=\frac{-i}{48}\Tr\left[\gamma_\nu\frac{\partial S^{-1}(p)}{\partial p_\nu}\right]_{p^2=\mu^2}.
\end{equation}
This is consistent with Ward Identities so that the renormalization constant in the RI scheme for the conserved vector
current is one. However on the lattice, it is not convenient to do derivatives with respect to the discretized momentum.

Following Ref.~\cite{Zhang:2005sca}, we shall use the renormalization of the axial-vector current to set the scale.
Since we can obtain the renormalization constant $Z_A^{WI}$ of the local axial vector current from Ward Identities, 
which equals to $Z_A^{RI}$ in the RI scheme, 
we can get $Z_q^{RI}$ from
\begin{equation}
Z_q^{RI}=Z_A^{WI}\frac{1}{12}\Tr\left[\Lambda_{A}(p)\Lambda_{A}^{tree}(p)^{-1}\right]_{p^2=\mu^2}.
\label{eq:zq_ri}
\end{equation}
Once we obtain $Z_q^{RI}$, we use Eq.(\ref{eq:ri_condition}) to get $Z_S$, $Z_P$ and $Z_V$
for the scalar, pseudoscalar and vector currents. At tree level, $\Lambda_{\mathcal{O}}^{tree}(p)=\Gamma$ for the quark bilinear operators.

The Green functions in Eq.(\ref{eq:ri_condition}) are not gauge invariant, therefore the calculation has to be
done in a fixed gauge, usually in the Landau gauge.

The massless overlap operator~\cite{Neuberger:1997fp} is defined as
\begin{equation}
D_{ov}  (\rho) =   1 + \gamma_5 \varepsilon (\gamma_5 D_{\rm w}(\rho)),
\end{equation}
where $\varepsilon$ is the matrix sign function and $D_{\rm w}(\rho)$ is the usual Wilson fermion operator, 
except with a negative mass parameter $- \rho = 1/2\kappa -4$ in which $\kappa_c < \kappa < 0.25$. 
We set $\kappa = 0.2$ in our calculation that corresponds to $\rho = 1.5$. The massive overlap Dirac operator is defined as
\begin{eqnarray}
D_m &=& \rho D_{ov} (\rho) + m\, (1 - \frac{D_{ov} (\rho)}{2}) \nonumber\\
       &=& \rho + \frac{m}{2} + (\rho - \frac{m}{2})\, \gamma_5\, \varepsilon (\gamma_5 D_w(\rho)).
\end{eqnarray}
To accommodate the $SU(3)$ chiral transformation, it is usually convenient to use the chirally regulated field
$\hat{\psi} = (1 - \frac{1}{2} D_{ov}) \psi$ in lieu of $\psi$ in the interpolation field and the currents.
This is equivalent to leaving the unmodified currents and instead adopting the effective propagator
%
\begin{equation}
G \equiv D_{eff}^{-1} \equiv (1 - \frac{D_{ov}}{2}) D^{-1}_m = \frac{1}{D_c + m},
\end{equation}
where $D_c = \frac{\rho D_{ov}}{1 - D_{ov}/2}$ is chiral, i.e. $\{\gamma_5, D_c\}=0$~\cite{Chiu:1998gp}.
With the good chiral properties of overlap fermions,
we should get $Z_S=Z_P$ and $Z_V=Z_A$. These relations are well satisfied within uncertainties by our numerical results as will be shown later.

\section{Numerical results}
\label{sec:num_results}
Our configurations are generated by the RBC-UKQCD collaboration using 2+1 flavor domain wall fermions\cite{Aoki:2010dy,Allton:2008pn}. The lattice sizes
are $24^3\times64$ and $32^3\times64$. On each lattice, there are three different light sea quark masses. On the $24^3\times64$ lattice
they are $m_l/m_s=0.005/0.04, 0.01/0.04$ and $0.02/0.04$ in lattice units. On the $32^3\times64$ lattice, $m_l/m_s=0.004/0.03, 0.006/0.03$ and $0.008/0.03$.
We employ one HYP smearing on the gauge fields~\cite{Li:2010pw} and then fix to the Landau gauge. The corresponding rotation matrices are saved.
Then the quark propagators in the Landau gauge are rotated from those already computed before the gauge fixing to save time by avoiding doing inversions.
The effects of smearing (one or only a few iterations) on observables go away in the continuum limit. 
Also, note that HYP smearing and gauge rotation on a configuration commute.
Thus the effects in vertex functions 
of doing smearing before or after gauge fixing, or not doing smearing at all differ by discretization effects at a fixed lattice spacing.
In Tab.~\ref{tab:nconfs}, we give the number of configurations used in this work on each data ensemble.
\begin{table}
\begin{center}
\caption{The number of configurations used in this work on the $24^3\times64$ and $32^3\times64$ lattices. The residual masses of DWF in lattice units $m_{res}$
are in the two-flavor chiral limit as given in Ref.~\cite{Aoki:2010dy}.}
\begin{tabular}{ccccc}
\hline\hline
label & $m_l/m_s$ & volume & $N_{conf}$ & $m_{res}$ \\
\hline
c005  & 0.005/0.04 & $24^3\times64$ & 92 & 0.003152(43) \\
c01  & 0.01/0.04 & $24^3\times64$ & 88  & \\
c02 & 0.02/0.04 & $24^3\times64$ & 138 & \\
\hline
f004 & 0.004/0.03 & $32^3\times64$ & 50 & 0.0006664(76) \\
f006 & 0.006/0.03 & $32^3\times64$ & 40 & \\
f008 & 0.008/0.03 & $32^3\times64$ & 50 & \\
\hline\hline
\end{tabular}
\label{tab:nconfs}
\end{center}
\end{table}
The overlap valence quark masses in lattice units are given in Tab.~\ref{tab:24_32}. 
The corresponding pion masses are from about 220 to 600 MeV.
\begin{table}
\begin{center}
\caption{Overlap valence quark masses in lattice units on the $24^3\times64$ and $32^3\times64$ lattices.}
\begin{tabular}{ccccccccc}
\hline\hline
$24^3\times64$  & 0.00620 & 0.00809 & 0.01020 & 0.01350 & 0.01720 & 0.02430 & 0.03650 & 0.04890 \\
\hline
$32^3\times64$ &   0.00460 & 0.00585 & 0.00677 & 0.00885  &   0.01290 & 0.01800 & 0.02400 & 0.03600 \\ 
\hline\hline
\end{tabular}
\label{tab:24_32}
\end{center}
\end{table}

We use anti-periodic boundary condition in the time direction and periodic boundary condition in the spacial directions. Therefore the momenta are
\begin{equation}
ap=(\frac{2\pi k_1}{L}, \frac{2\pi k_2}{L}, \frac{2\pi k_3}{L}, \frac{(2k_4+1)\pi}{T}),
\end{equation}
where $k_\mu=-6,-5,...,6$ on the $L=24$ lattice and $k_i=-6,-7,...,6$, $k_4=-5,-1,...,6$ on the $L=32$ lattice. To reduce the effects of Lorentz
non-invariant discretization errors, we only use the momenta which satisfy the condition
\begin{equation}
\frac{p^{[4]}}{(p^2)^2}<0.32,\quad\mbox{where } p^{[n]}=\sum_{\mu=1}^4 p_\mu^{n},\quad p^2=\sum_\mu p_\mu^2.
\label{eq:p4p22}
\end{equation}
In other words, only those momenta pointing close to the diagonal direction are used.
However as the statistical error decreases (for example, by using momentum sources~\cite{Gockeler:1998ye}), 
the effects proportional to $a^2p^{[4]}/p^2$
can be seen.
To use all momenta and systematically remove the 
hypercubic effects, one can follow the method used in Refs.~\cite{deSoto:2007ht,Blossier:2010vt}.
Another way is to follow Ref.~\cite{Arthur:2010ht}.
One can also use perturbative calculations to subtract and suppress those effects as, for example, in Ref.~\cite{Constantinou:2013ada}.

In our calculation, we require same $p_4$, $p^{[4]}$ and $p^{[6]}$ when averaging momentum modes with a same $p^2$.
Therefore we can estimate the $\mathcal{O}(a^2p^{[4]}/p^2)$ lattice artifacts (ignoring higher terms).
As we will show later, those effects are not small in $Z_S$.
But because the condition in Eq.(\ref{eq:p4p22}) is used, the $\mathcal{O}(a^2p^{[4]}/p^2)$ effects can be absorbed into
a simple $\mathcal{O}(a^2p^2)$ term within our statistical uncertainty.

We use point source propagators in the Landau gauge to evaluate all the necessary Green functions and vertex functions. 
Momentum sources~\cite{Gockeler:1998ye}
can be used to improve the signal-to-noise ratio. But for each momentum one inversion is needed, which is expensive for overlap fermions.
Thus we use the point source propagators which can be projected to many momenta. The statistical errors of
our final results are from Jackknife processes.

\subsection{Renormalization of the axial vector current from Ward identity}
\label{sec:za_wi}
The renormalization constant $Z_A$ can be obtained from the axial Ward identity
\begin{equation}
Z_A\partial_\mu A_\mu=2Z_m m_q Z_P P,
\label{eq:ZA_WI}
\end{equation}
where $A_\mu$ and $P$ are the local axial vector current and the pseudoscalar density and $Z_m$ is the quark mass
renormalization constant with the renormalized mass $m_R=Z_m m_q$. Since $Z_m=Z_P^{-1}$ for overlap fermions,
one can find $Z_A$ by considering the matrix elements of the both sides of Eq.(\ref{eq:ZA_WI}) between the vacuum and a pion
\begin{equation}
Z_A\partial_\mu\langle 0| A_\mu |\pi\rangle=2m_q\langle 0|P|\pi\rangle.
\end{equation}
If the pion is at rest, then from the above equation one gets
\begin{equation}
Z_A=\frac{2m_q\langle0|P|\pi\rangle}{m_\pi\langle 0| A_4 |\pi\rangle},
\end{equation}
where $A_4=\bar\psi\gamma_4\gamma_5\hat\psi$ and $P=\bar\psi\gamma_5\hat\psi$. To obtain the matrix elements, we compute 2-point correlators
\begin{equation}
G_{PP}(\vec p=0,t)=\sum_{\vec x}\langle0|P(x)P^\dagger(0)|0\rangle,
\end{equation}
and
\begin{equation}
G_{A_4P}(\vec p=0,t)=\sum_{\vec x}\langle0|A_4(x)P^\dagger(0)|0\rangle.
\end{equation}
When the time $t$ is large, the contribution from the pion dominates in both correlators. Then one has
\begin{equation}
Z_A^{WI}=\lim_{m_q\rightarrow0,t\rightarrow\infty}\frac{2m_qG_{PP}(\vec p=0,t)}{m_\pi G_{A_4P}(\vec p=0,t)}.
\label{eq:Za_WI_2pt}
\end{equation}

\begin{figure}
\begin{center}
\includegraphics[height=2.6in,width=0.49\textwidth]{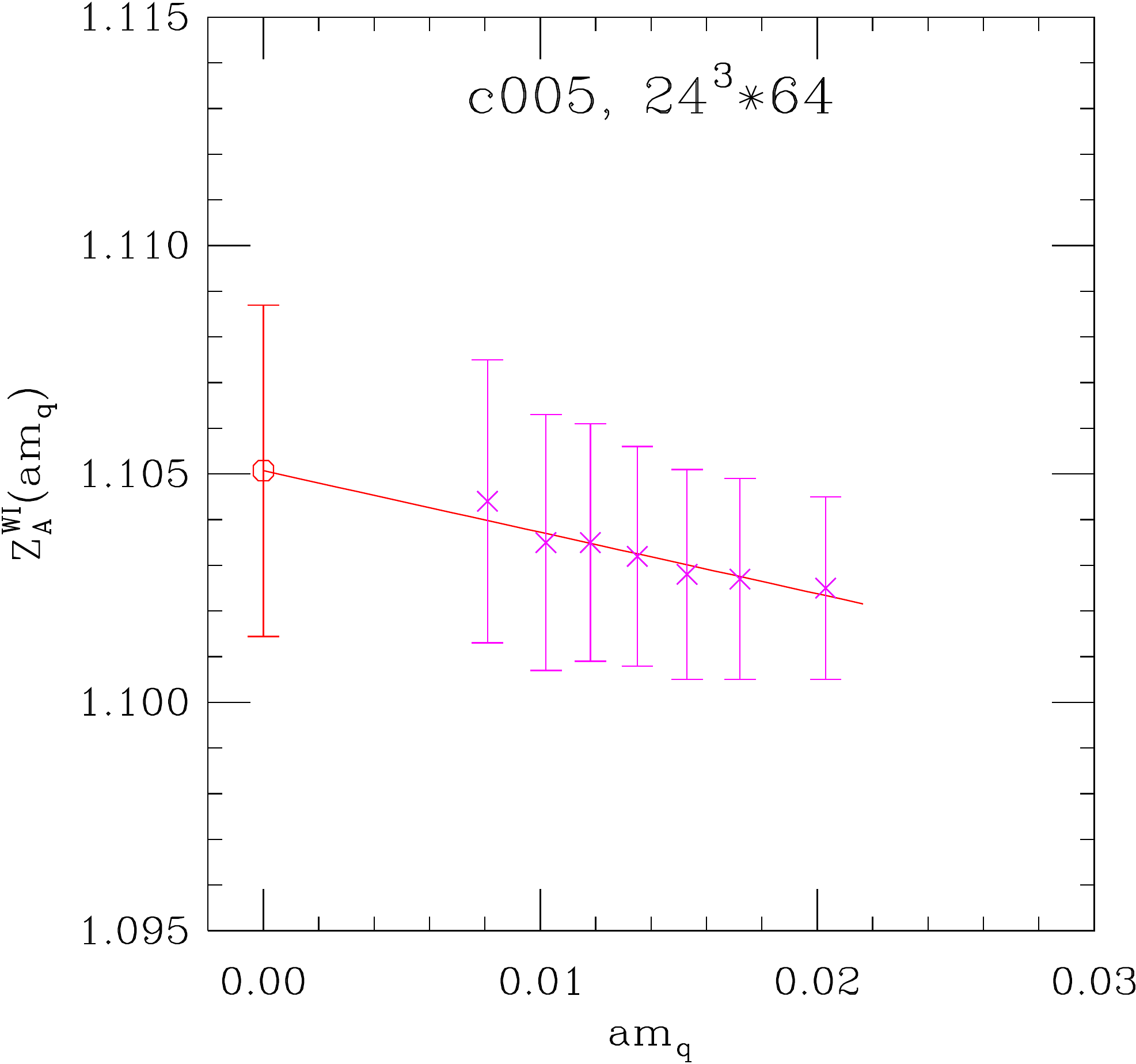}
\includegraphics[height=2.6in,width=0.49\textwidth]{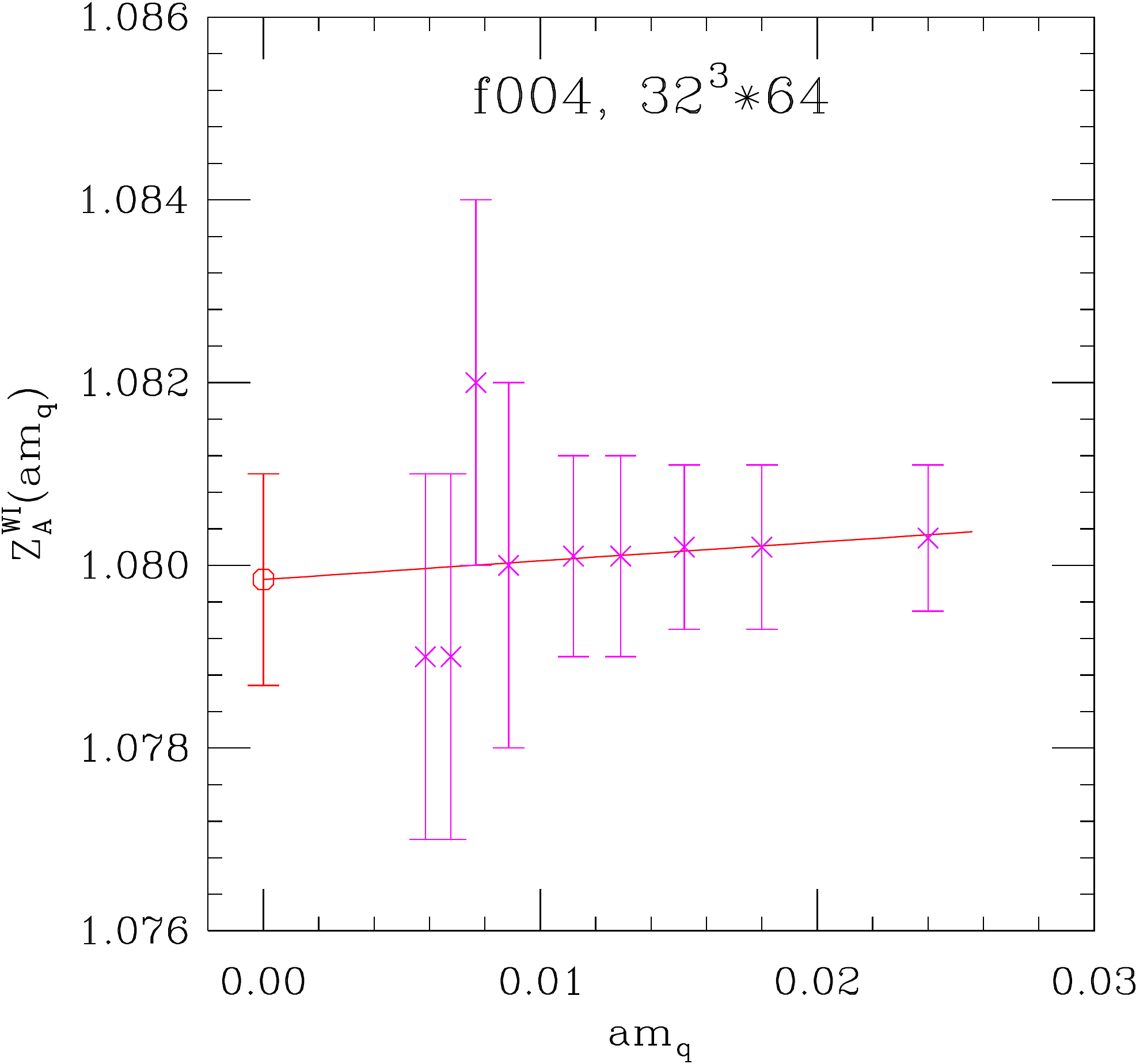}
\end{center}
\caption{Examples of $Z_A^{WI}(am_q)$ against valence quark masses. The left graph is for the $L=24$
lattice with sea quark masses $m_l/m_s=0.005/0.04$. The right one for the $L=32$ lattice with $m_l/m_s=0.004/0.03$.}
\label{fig:Za_WI}
\end{figure}

In Fig.~\ref{fig:Za_WI} we show examples of $Z_A^{WI}$ obtained from Eq.(\ref{eq:Za_WI_2pt}) before taking the valence quark massless limit 
(denoted as $Z_A^{WI}(am_q)$).
To take the limit $m_q\rightarrow0$, we fit the data to~\cite{Zhang:2005sca}
\begin{equation}
Z_A^{WI}(am_q)=Z_A^{WI}(1+b_A am_q).
\end{equation}
After taking the valence quark massless limit, we get the results of $Z_A^{WI}$ as given in Tab.~\ref{tab:Za_WI_24_32}.
\begin{table}
\begin{center}
\caption{$Z_A^{WI}$ on the $24^3\times64$ and $32^3\times64$ lattices.}
\begin{tabular}{cccccc}
\hline\hline
$24^3\times64$ & $m_l/m_s$ &  0.02/0.04 & 0.01/0.04 & 0.005/0.04 &  $m_l+m_{res}=0$ \\
 & $Z_A^{WI}$ &   1.101(4) & 1.115(6) & 1.105(4) &  1.111(6) \\
\hline
$32^3\times64$ & $m_l/m_s$ & 0.008/0.03 & 0.006/0.03 &  0.004/0.03 & $m_l+m_{res}=0$ \\
 & $Z_A^{WI}$ & 1.075(1) & 1.079(1) & 1.080(1) & 1.086(2) \\
\hline\hline
\end{tabular}
\label{tab:Za_WI_24_32}
\end{center}
\end{table}
In the last column of Tab.~\ref{tab:Za_WI_24_32}, the results at the light sea quark massless limit are obtained by a linear extrapolation in $m_l+m_{res}$,
where $m_{res}$ is given in Tab.~\ref{tab:nconfs}.

\subsection{Analysis of the quark propagator}
\begin{figure}
\begin{center}
\includegraphics[height=2.6in,width=0.49\textwidth]{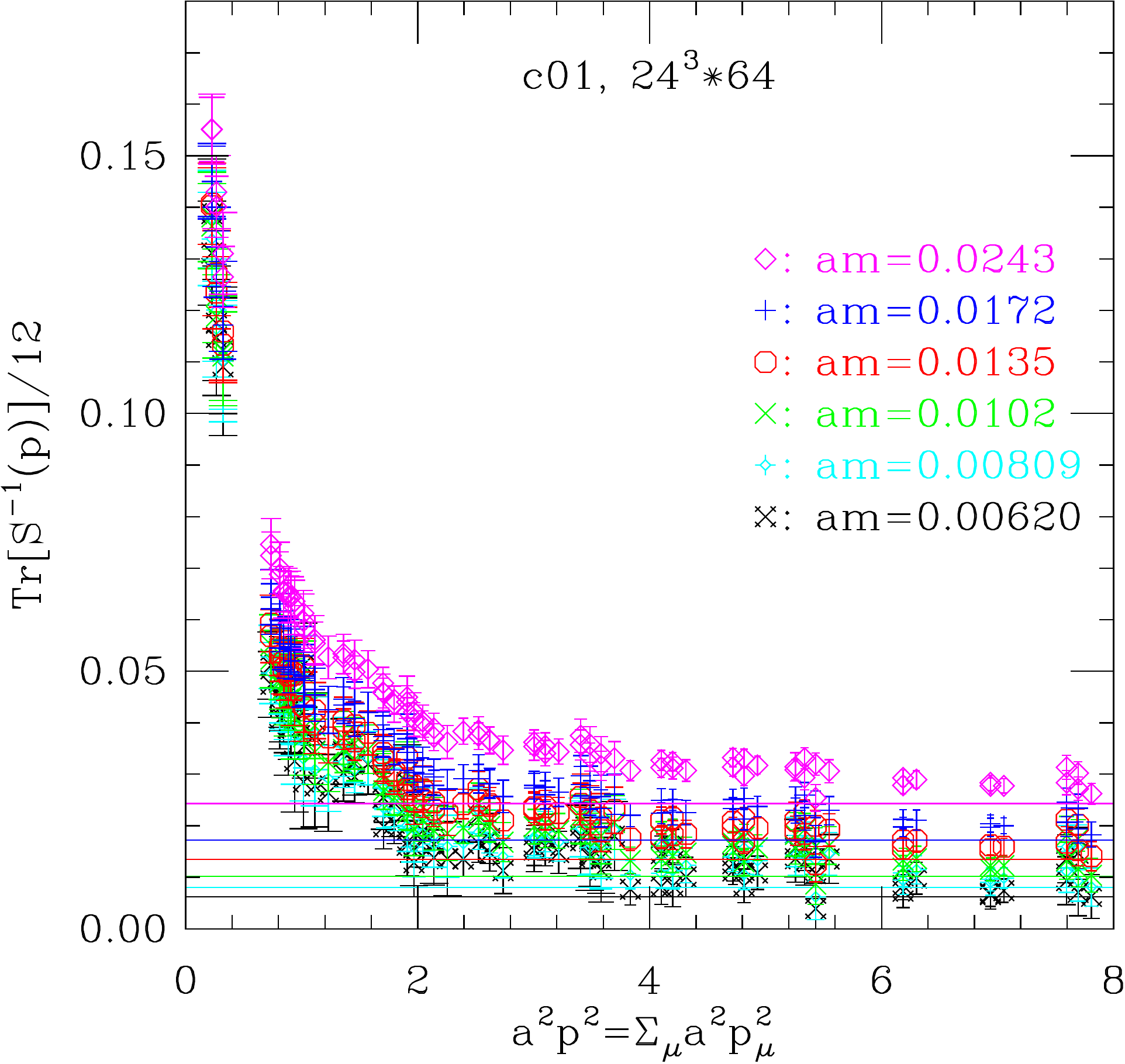}
\includegraphics[height=2.6in,width=0.49\textwidth]{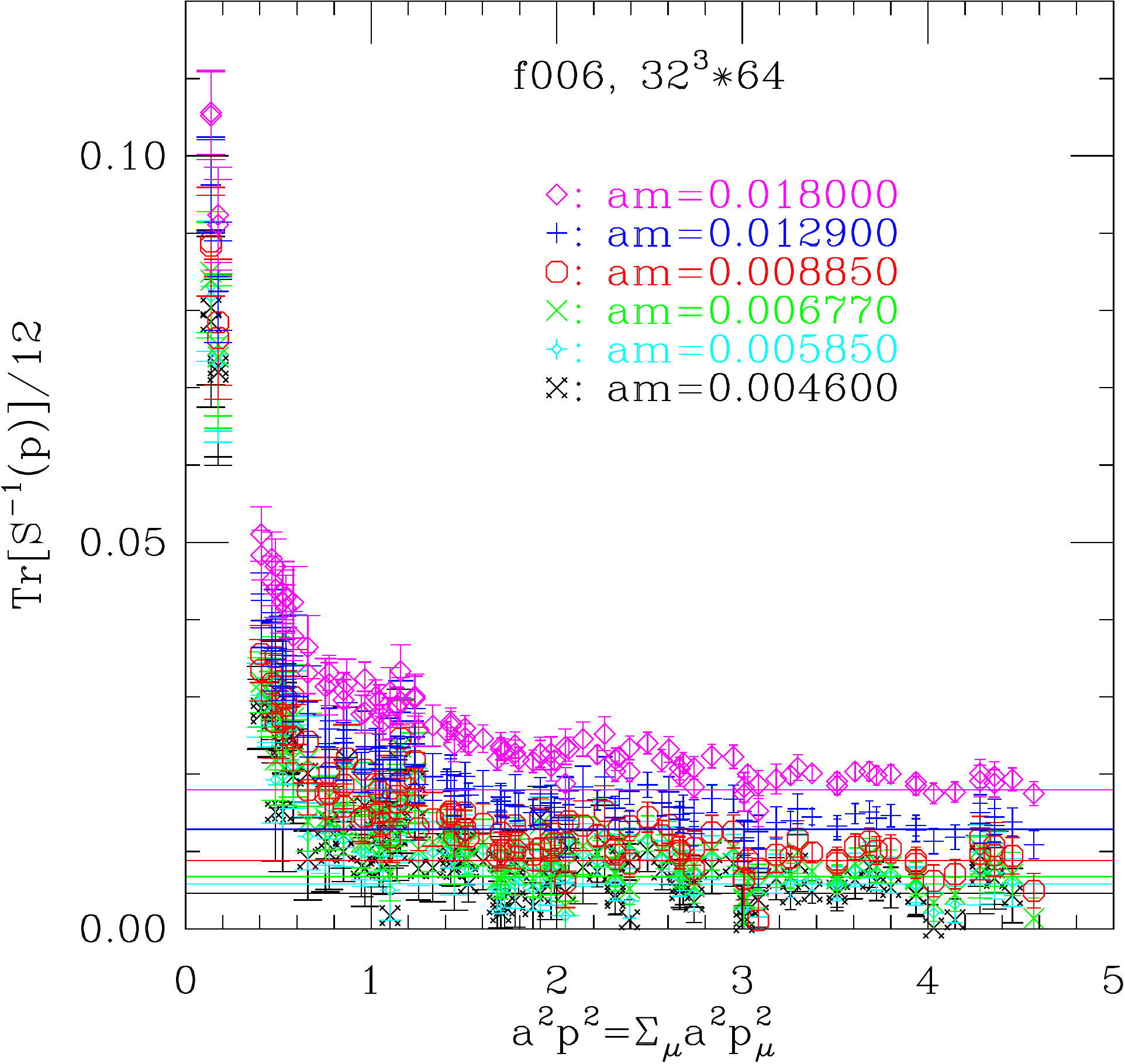}
\end{center}
\caption{Examples of $\Tr(S^{-1}(p))/12$ as functions of the momentum scale for different bare valence quark masses. The left graph is for the $L=24$
lattice with sea quark masses $m_l/m_s=0.01/0.04$. The right one for the $L=32$ lattice with $m_l/m_s=0.006/0.03$. The horizontal lines are the
positions of the bare quark masses.}
\label{fig:scalarm}
\end{figure}

At large momentum, because of asymptotic freedom the quark propagator $S(p)$ goes back to the free quark propagator. In Fig.~\ref{fig:scalarm} we show examples of
$\Tr(S^{-1}(p))/12$ as functions of the momentum scale for different bare valence quark masses. As expected, $\Tr(S^{-1}(p))/12$ goes to the bare quark mass value
as the momentum scale increases. The two graphs in Fig.~\ref{fig:scalarm} are for data ensemble c01 and f006 respectively.
The results from other ensembles are similar.

\begin{figure}
\begin{center}
\includegraphics[height=2.6in,width=0.49\textwidth]{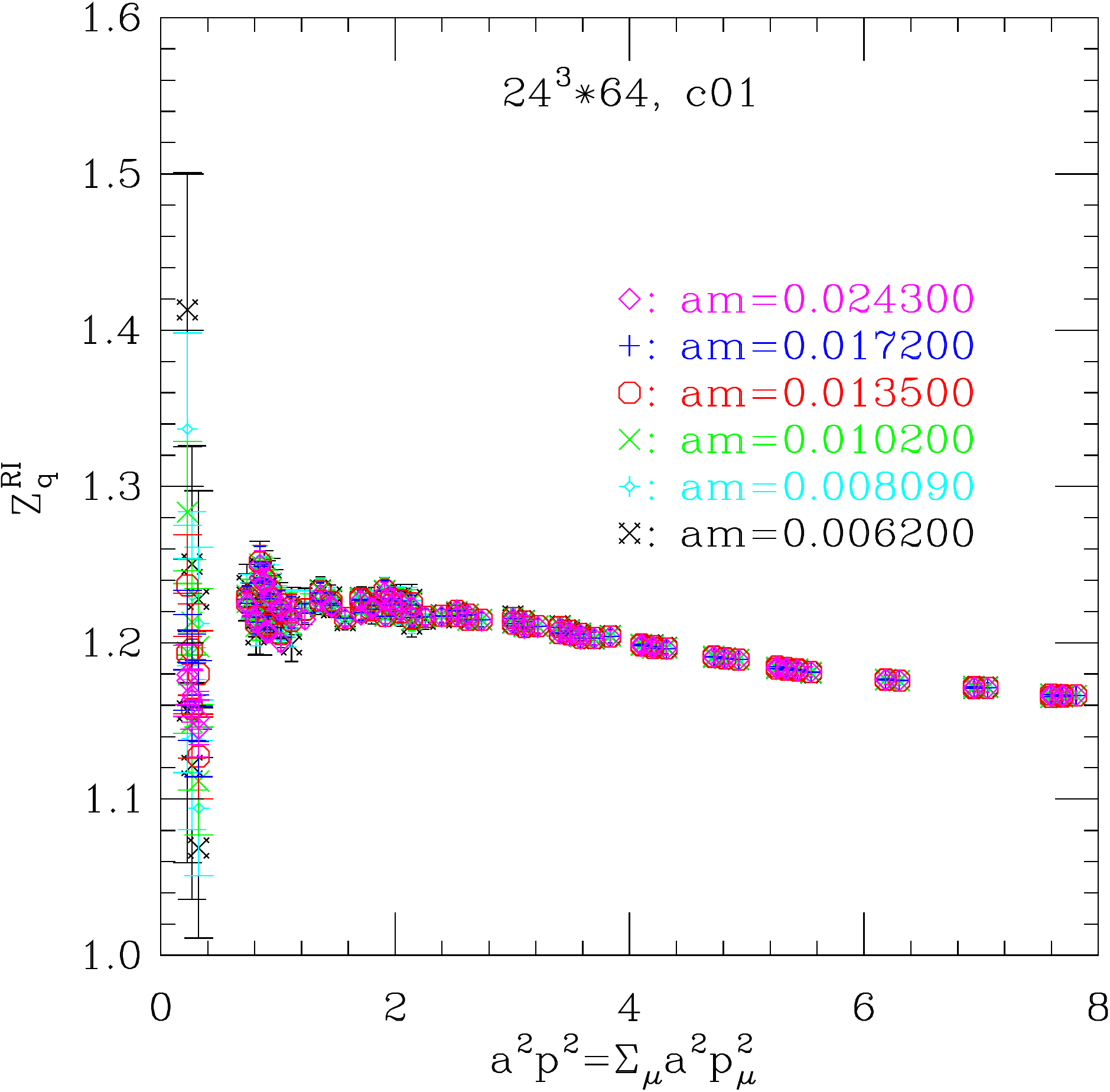}
\includegraphics[height=2.6in,width=0.49\textwidth]{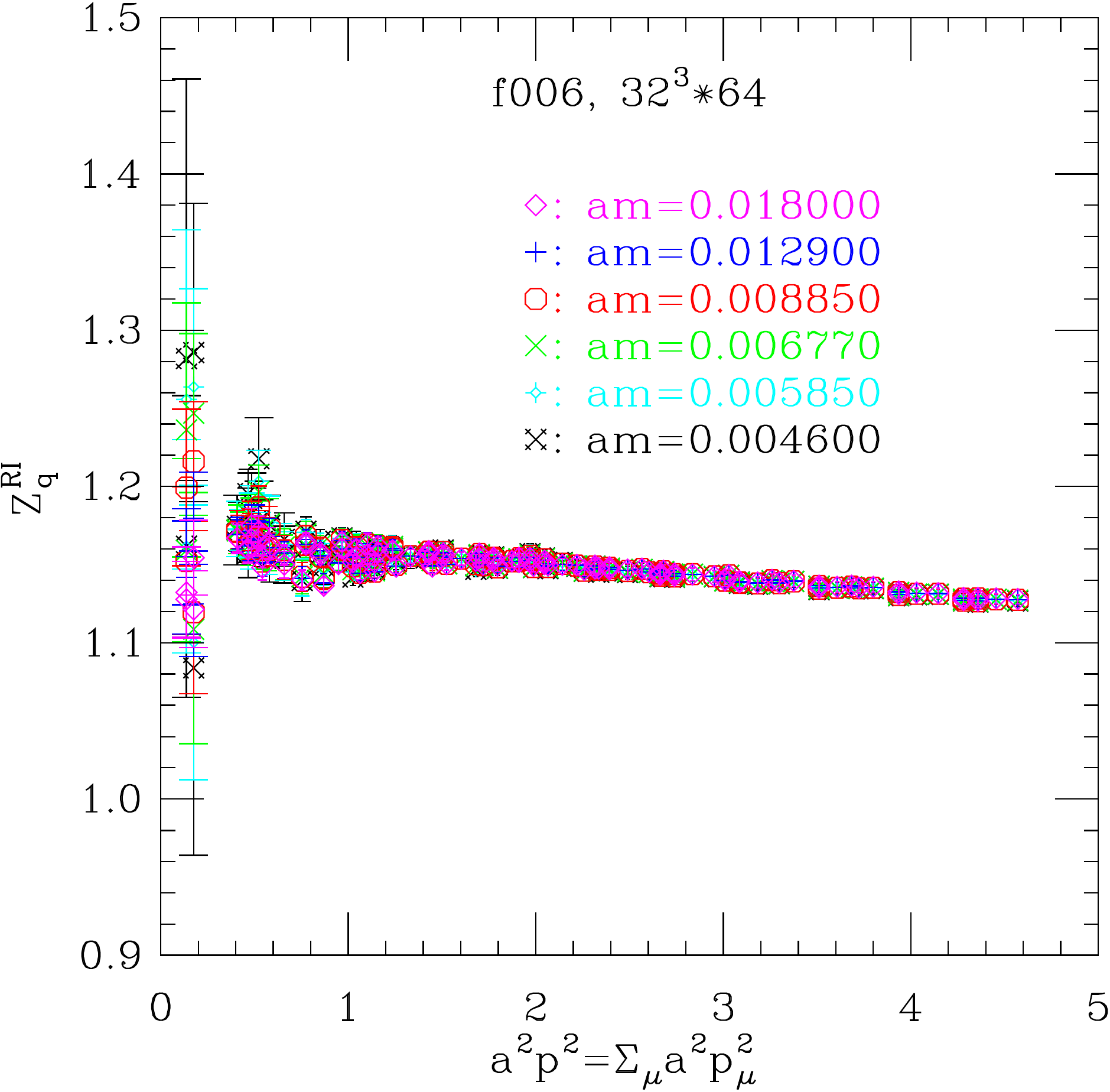}
\end{center}
\caption{Examples of $Z_q^{RI}$ as functions of the momentum scale for different valence quark masses. The left graph is for the $L=24$
lattice with sea quark masses $m_l/m_s=0.01/0.04$. The right one for the $L=32$ lattice with $m_l/m_s=0.006/0.03$.}
\label{fig:zq_ri}
\end{figure}

Fig.~\ref{fig:zq_ri} shows examples of the quark field renormalization constants $Z_q^{RI}$ as functions of the momentum scale for different
valence quark masses.
$Z_q^{RI}$ is computed from Eq.(\ref{eq:zq_ri}). As we can see, the quark mass dependence of $Z_q^{RI}$ is quite small on both the $L=24$ and 32
lattices. The symbols in Fig.~\ref{fig:zq_ri} are on top of each other except at very small $a^2p^2$.

In Landau gauge, the anomalous dimension of $Z_q$ is zero at 1-loop. This is why in Fig.~\ref{fig:zq_ri} the behavior of $Z_q$ is quite flat
up to $\mathcal{O}(a^2p^2)$ discretization errors.

\subsection{Scalar density}
\label{sec:zs_ri}
\begin{figure}
\begin{center}
\includegraphics[height=2.6in,width=0.49\textwidth]{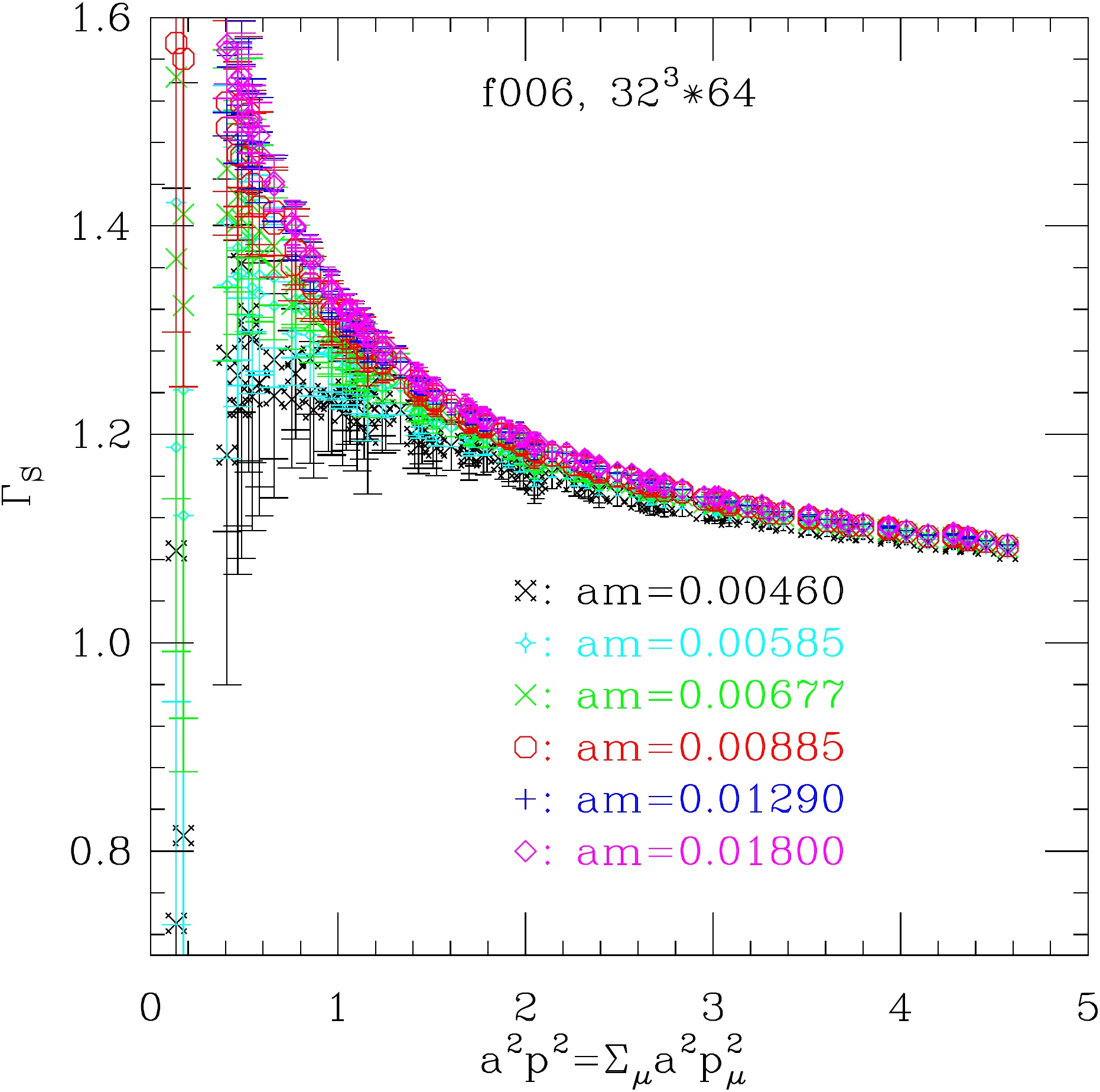}
\includegraphics[height=2.6in,width=0.49\textwidth]{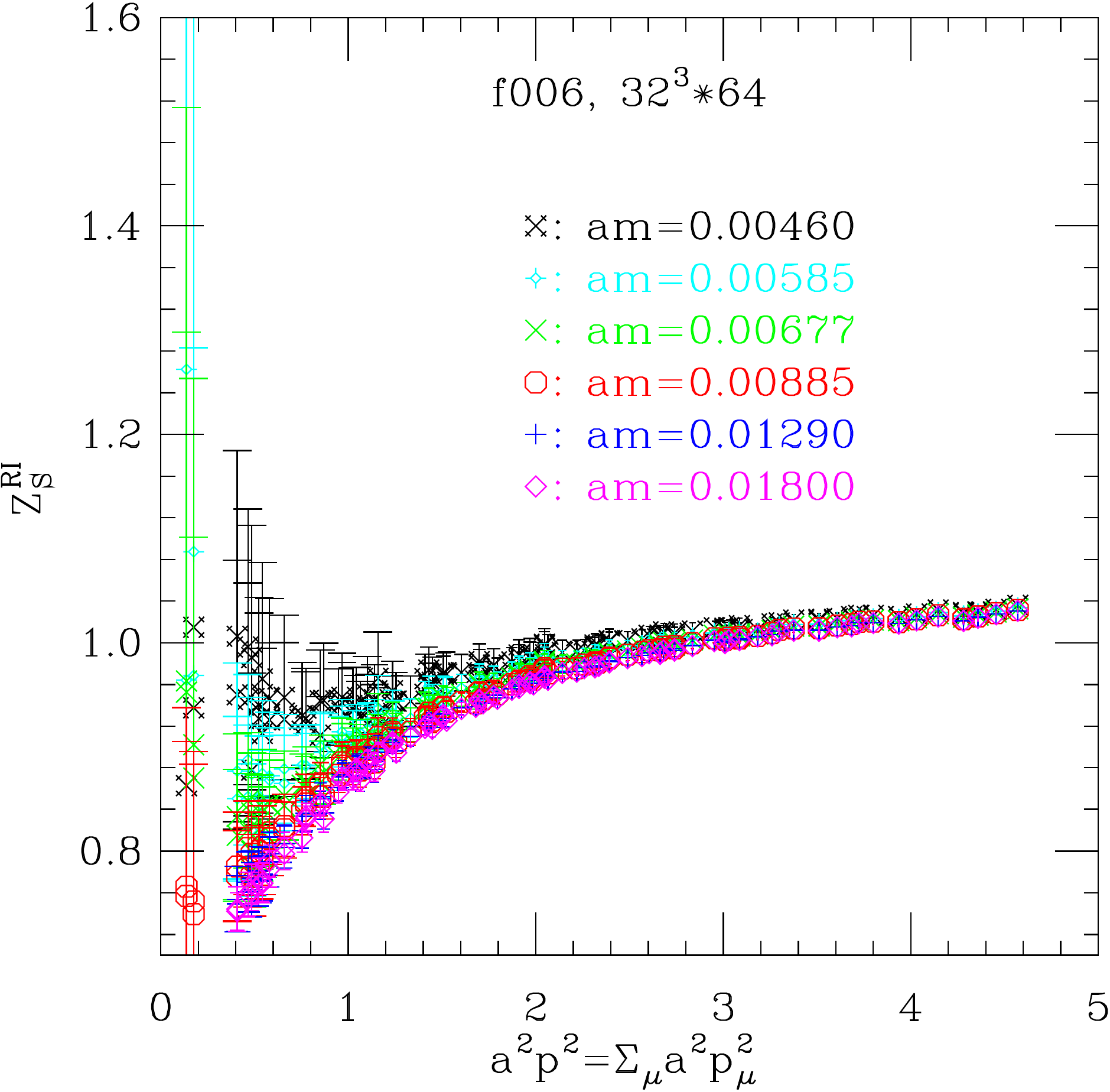}
\end{center}
\caption{Examples of the projected vertex function $\Gamma_S$ and $Z_S^{RI}$ as functions of the momentum scale for ensemble f006.}
\label{fig:zs_ri}
\end{figure}

After obtaining $Z_q^{RI}$, one can now get $Z_S^{RI}$ from Eq.(\ref{eq:ri_condition}). The projected vertex function $\Gamma_S$ (defined in
Eq.(\ref{eq:GammaS})) and $Z_S^{RI}$ as functions of the momentum scale for different valence quark masses
on ensemble f006 are shown in Fig.~\ref{fig:zs_ri}.

\begin{figure}
\begin{center}
\includegraphics[height=2.6in,width=0.49\textwidth]{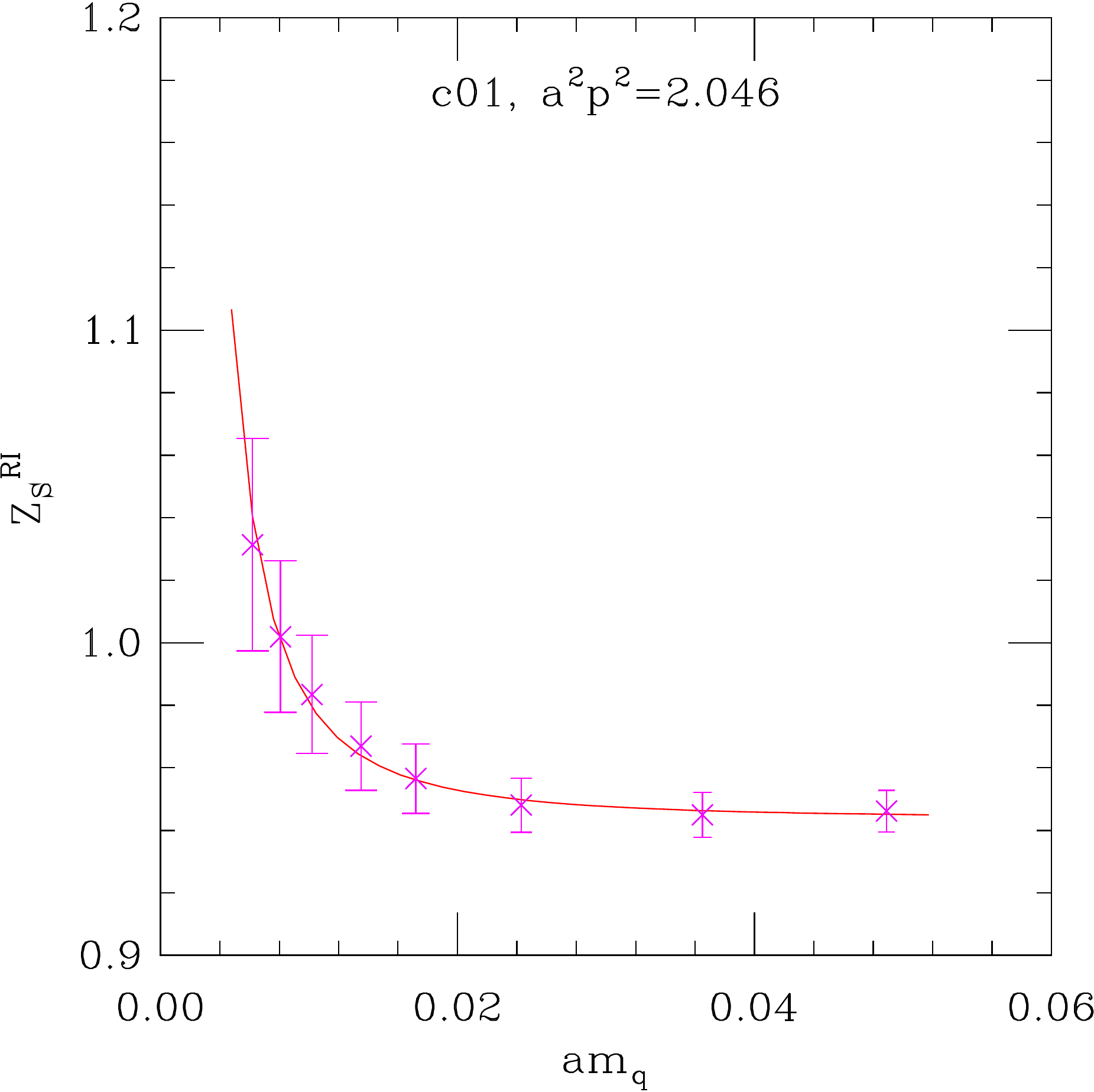}
\includegraphics[height=2.6in,width=0.49\textwidth]{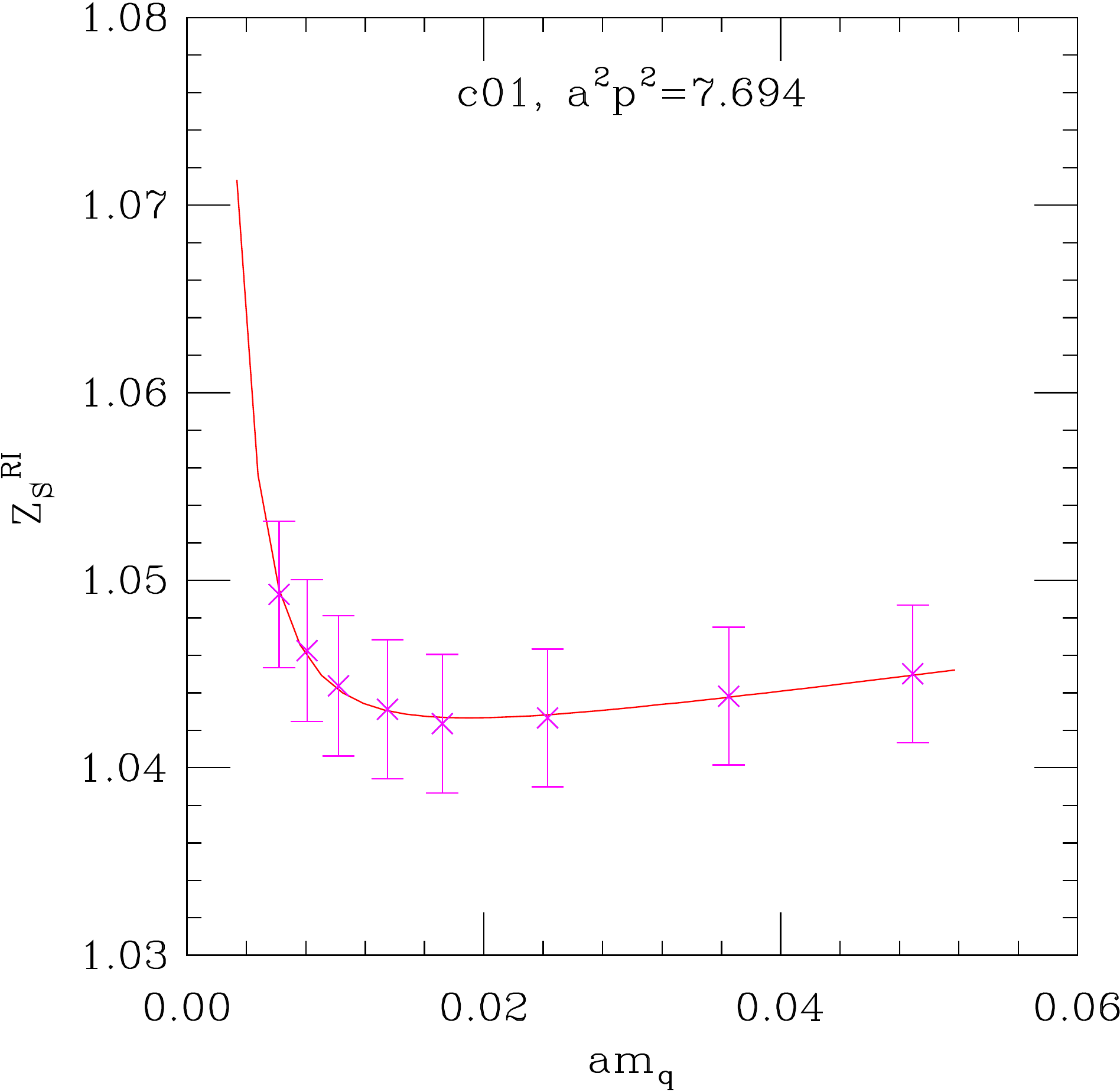}
\end{center}
\caption{$Z_S^{RI}$ as a function of the valence quark mass at two momentum scales for ensemble c01. The curves are fits to Eq.(\ref{eq:zs_fit}).}
\label{fig:zs_amq}
\end{figure}

Fig.~\ref{fig:zs_amq} shows $Z_S^{RI}$ as a function of the valence quark mass at different momentum for ensemble c01.
Apparently, the dependence on
$am_q$ is not linear. Thus to go to the chiral limit, we use
\begin{equation}
Z_S=\frac{A_s}{(am_q)^2}+B_s+C_s(am_q)
\label{eq:zs_fit}
\end{equation}
to fit our data and take $B_s$ as the chiral limit value of $Z_S$. This fit function is inspired from Refs.~\cite{Blum:2001sr,Aoki:2007xm}. 
The double pole term in the above equation comes from the topological zero modes of the overlap fermions.
In a calculation of $Z_S$ in the RI' scheme~\cite{DeGrand:2005af},
the curving up of $Z_S$ at small valence quark mass
is suppressed when the zero modes are subtracted from the quark propagator.

\begin{figure}
\begin{center}
\includegraphics[width=0.4\textwidth]{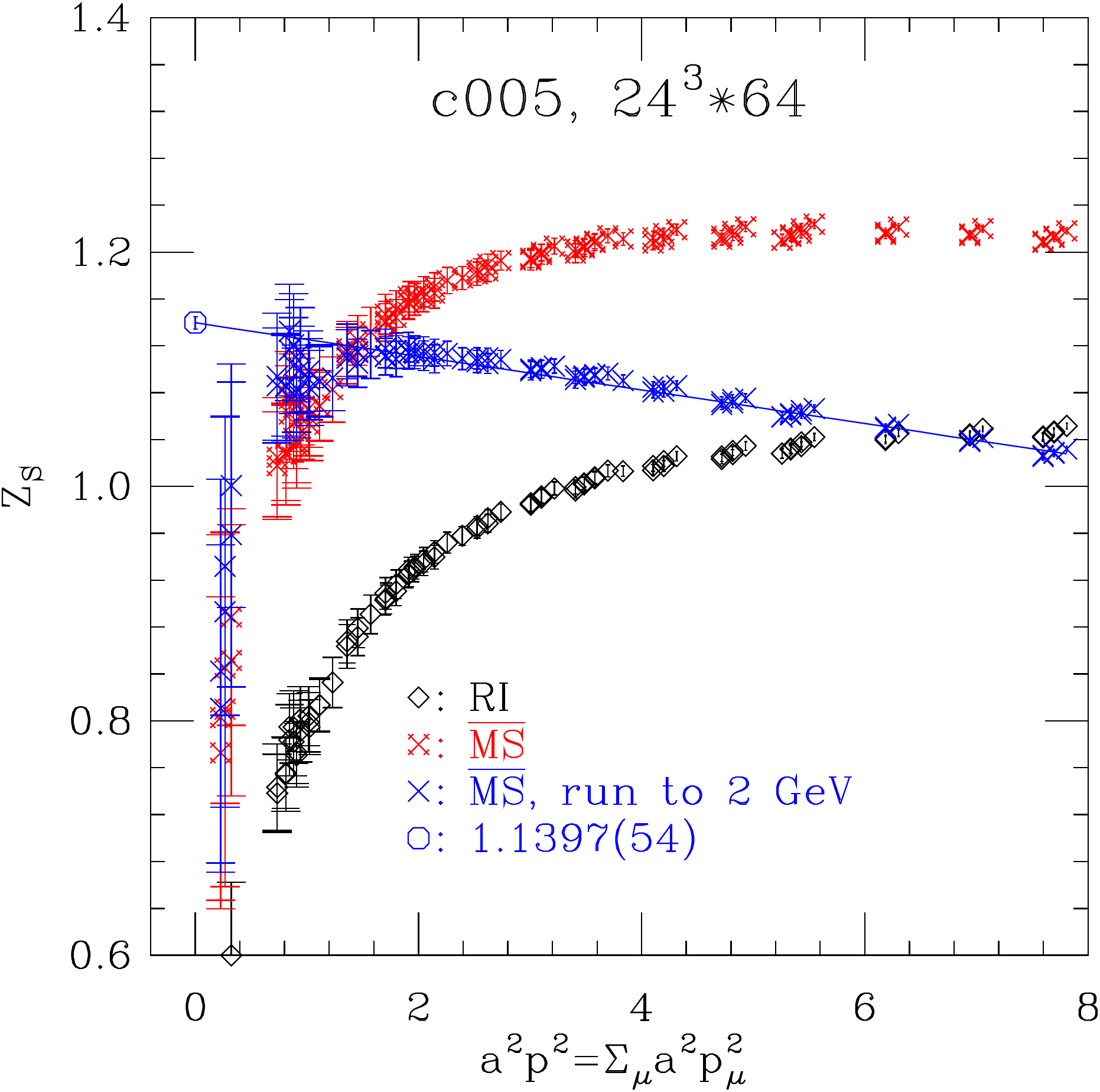}
\includegraphics[width=0.4\textwidth]{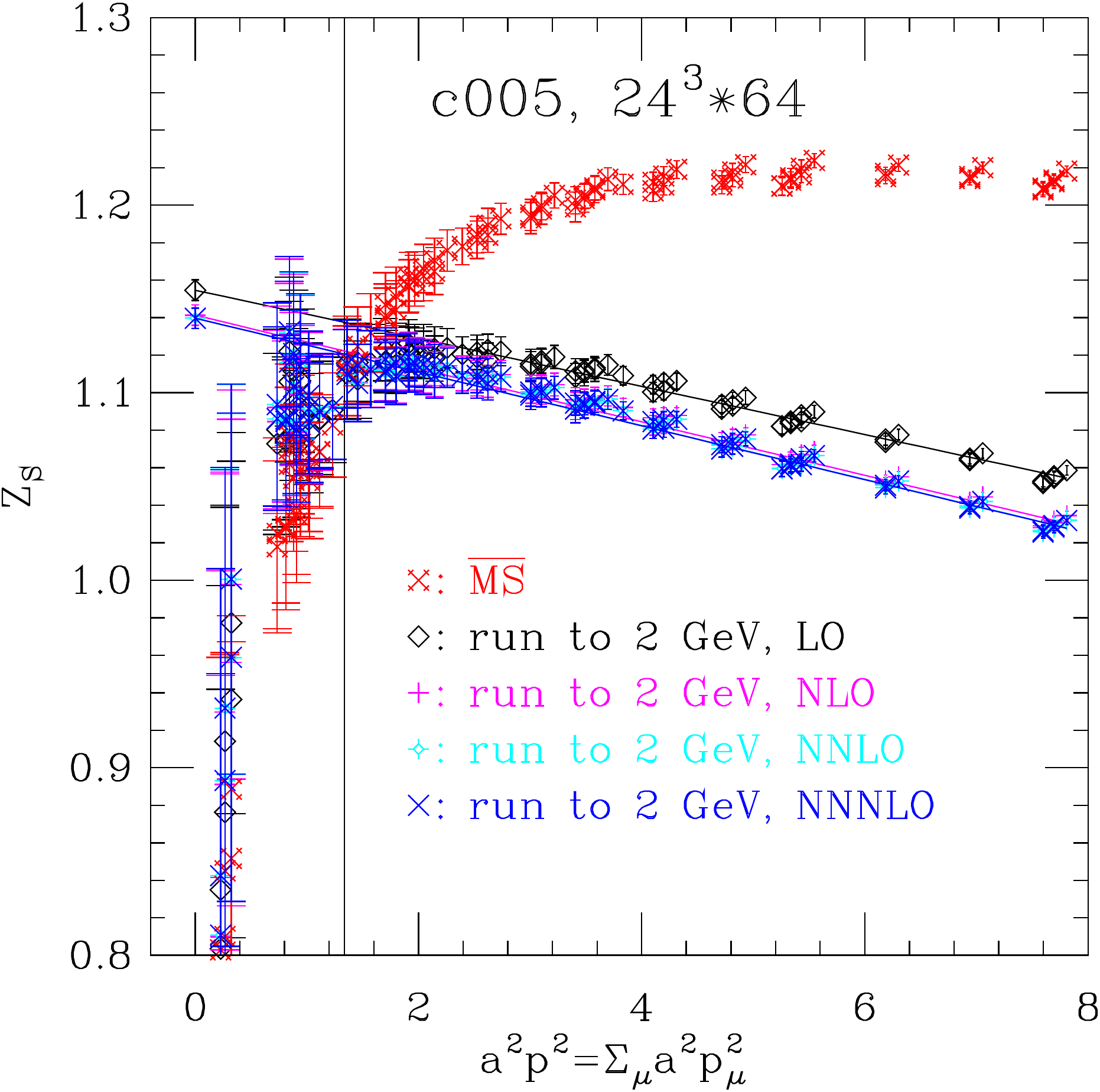}
\end{center}
\caption{The conversion and running of $Z_S$ in the valence quark massless limit on ensemble c005. Left panel: The black diamonds are the values in
the RI scheme. The red fancy crosses are those in the $\msbar$ scheme. The blue crosses are the results evolved to 2 GeV in the
$\msbar$ scheme as a function of the initial renormalization scale. Right panel: Comparison of the different orders of perturbative running 
in the $\msbar$ scheme. The vertical line indicates $p=2$ GeV.}
\label{fig:zs_msbar}
\end{figure}

The fits of the data to Eq.(\ref{eq:zs_fit}) have small $\chi^2/$dof at all momentum scales. Examples are shown in Fig.~\ref{fig:zs_amq}.
The results of $Z_S^{RI}$ in the valence quark massless limit as a function of the momentum scale for ensemble c005 are shown 
by the black diamonds in the left panel of Fig.~\ref{fig:zs_msbar}.

Then one can use conversion ratios calculated in continuum perturbation theory to convert $Z_S^{RI}$ into the $\msbar$ scheme.
In the quark massless limit, in Landau gauge and to three loops, 
the conversion ratio for $Z_S$ and $Z_P$ is~\cite{Franco:1998bm,Chetyrkin:1999pq}
\begin{eqnarray}
\frac{Z_S^{\rm RI}}{Z_S^{\overline{\rm MS}}}&=&
\frac{Z_P^{\rm RI}}{Z_P^{\overline{\rm MS}}}
=1-\frac{16}{3}\frac{\alpha_s}{4\pi}+\left(-\frac{1990}{9}
+\frac{89n_f}{9}+\frac{152\zeta_3}{3}\right)
\left(\frac{\alpha_s}{4\pi}\right)^2
\nonumber \\
&&+\left(-\frac{6663911}{648}+\frac{236650n_f}{243}
-\frac{8918n_f^2}{729}+\frac{408007\zeta_3}{108}\right.
\label{eq:ratio_zs_zp}\\
&&-\left.\frac{4936\zeta_3n_f}{27}
-\frac{32\zeta_3n_f^2}{27}
+\frac{80\zeta_4n_f}{3}
-\frac{2960\zeta_5}{9}\right)\left(\frac{\alpha_s}{4\pi}\right)^3
+O(\alpha_s^4),\nonumber
\end{eqnarray}
where $n_f$ is the number of flavors and $\zeta_n$ is the Riemann 
zeta function evaluated at $n$.

The value of $\alpha_s(\mu)$ is obtained by using its perturbative running to four loops~\cite{Alekseev:2002zn}.
The $\beta$-function in the $\msbar$ scheme to 4-loops is given in Ref.~\cite{vanRitbergen:1997va}.
We take the value $\Lambda_{QCD}^\msbar=339(10)$ MeV for three flavors in the $\msbar$ scheme~\cite{Beringer:1900zz} to evaluate Eq.(\ref{eq:ratio_zs_zp}) numerically.
For example, the strong coupling constant at 2 GeV is $\alpha_s^{\msbar}(2$ GeV$)=0.2787$.
The $\msbar$ value $Z_S^{\msbar}$ as a function of the scale $a^2p^2$ are shown by the red fancy crosses
in the left graph of Fig.~\ref{fig:zs_msbar}.

To obtain $Z_S^{\msbar}(2$ GeV$)$, we first use the anomalous dimension to four loops to
evolve $Z_S^\msbar(a^2p^2)$ at the initial renormalization scale $ap$ to 2 GeV (inverse lattice spacings $1/a=1.73$ GeV and 2.28 GeV are used
respectively).
Since $Z_S=Z_m^{-1}$, we can use the mass anomalous dimension given in Ref.~\cite{Chetyrkin:1999pq}
for the perturbative running.
The blue crosses in the left graph of Fig.~\ref{fig:zs_msbar} show $Z_S^\msbar(2$ GeV$; a^2p^2)$, which are the 4-loop running results from the 
initial renormalization scale $ap$ to the scale 2 GeV. $Z_S^\msbar(2$ GeV$; a^2p^2)$ would lie on a horizontal line at large
$a^2p^2$ if there were no discretization errors (and if the truncation error of the conversion ratio is small).

The solid blue line in the left panel of Fig.~\ref{fig:zs_msbar} is a linear fit to the blue crosses with $a^2p^2>5$. This is to reduce 
$\mathcal{O}(a^2p^2)$ discretization errors.
After the extrapolation we obtain $Z_S^\msbar (2$ GeV$)=1.1397(54)$ for c005, where the error is only statistical. If we use the blue crosses with
$a^2p^2>4$ to do the extrapolation, then we find $Z_S^\msbar (2$ GeV$)=1.1451(34)$. The two numbers are in agreement at one sigma.
The difference introduced by the different range of $a^2p^2$ will be included in the systematic errors of our final results.

\begin{figure}
\begin{center}
\includegraphics[width=0.4\textwidth]{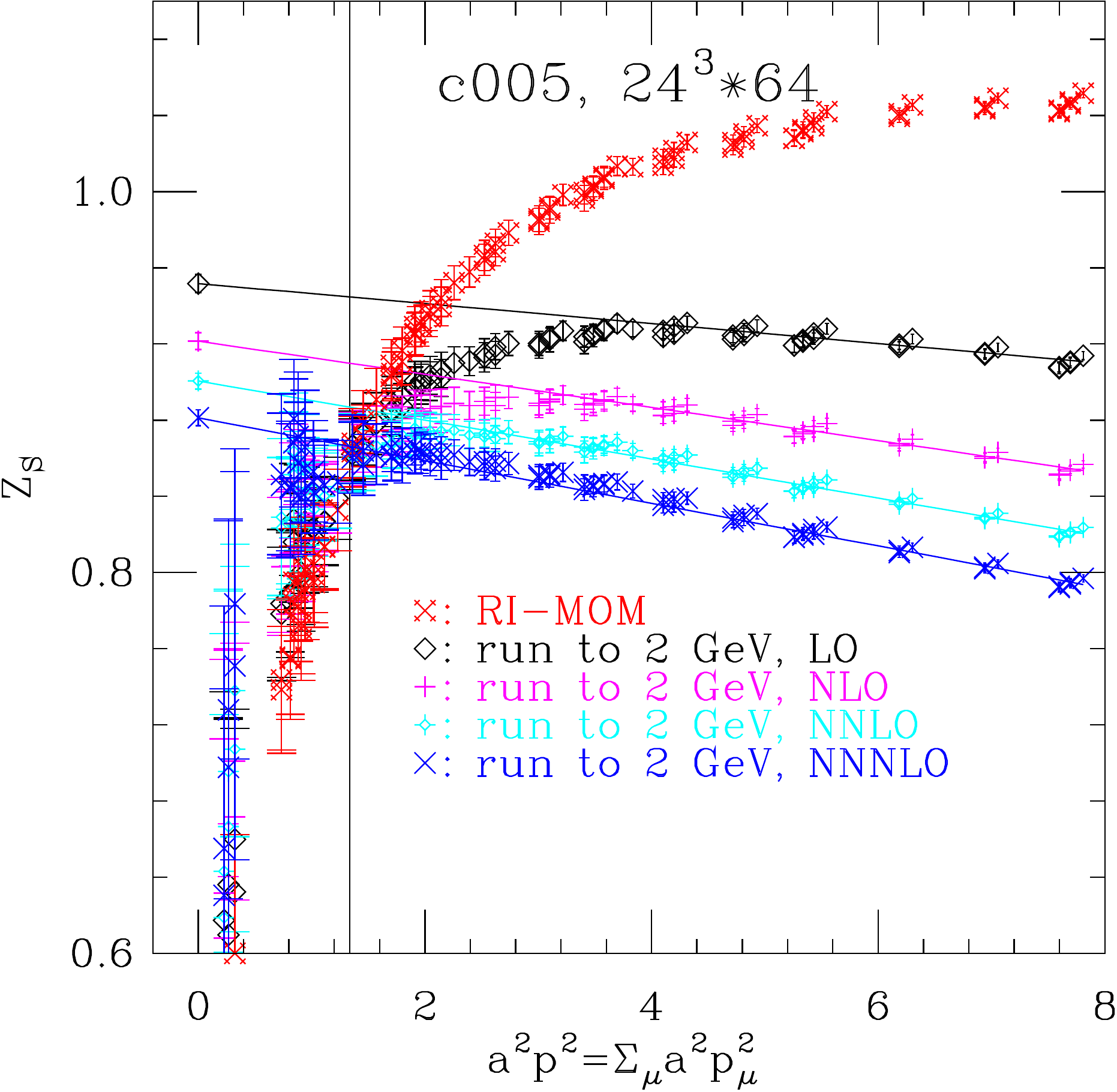}
\end{center}
\caption{Comparison of the different orders of perturbative running in the RI-MOM scheme. The vertical line indicates $p=2$ GeV.}
\label{fig:zs_running}
\end{figure}

In the right panel of Fig.~\ref{fig:zs_msbar}, we compare the different orders of perturbative running in the $\msbar$ scheme. 
As we can see, the truncation error is quite small after 2-loops. 
Only the 1-loop running results do not agree with the 4-loop (NNNLO) running results. This is in contrast with the 
truncation error of running in the RI-MOM scheme, which we show in Fig.~\ref{fig:zs_running}.
The perturbative truncation error for the running of $Z_S^{\rm RI}$ is large even with 4-loops: The 3-loop and 4-loop
results are different from each other. Similar behavior was also shown in Ref.~\cite{Aoki:2010yq}. 
With the much better running behavior in the $\msbar$ scheme, it is preferred not to do the 
perturbative running of $Z_S$ in the RI-MOM scheme. Nevertheless, if we use the $a^2p^2$ extrapolated
result $Z_S^{\rm RI}(2$ GeV$)=0.8812(41)$ after the 4-loop running and the conversion ratio 1.289614 from Eq.(\ref{eq:ratio_zs_zp}) 
at 2 GeV from the RI to the $\msbar$ scheme, we obtain $Z_S^\msbar(2$ GeV$)=1.1364(53)$. This is in agreement with the above 1.1397(54).

\begin{figure}
\begin{center}
\includegraphics[width=0.4\textwidth]{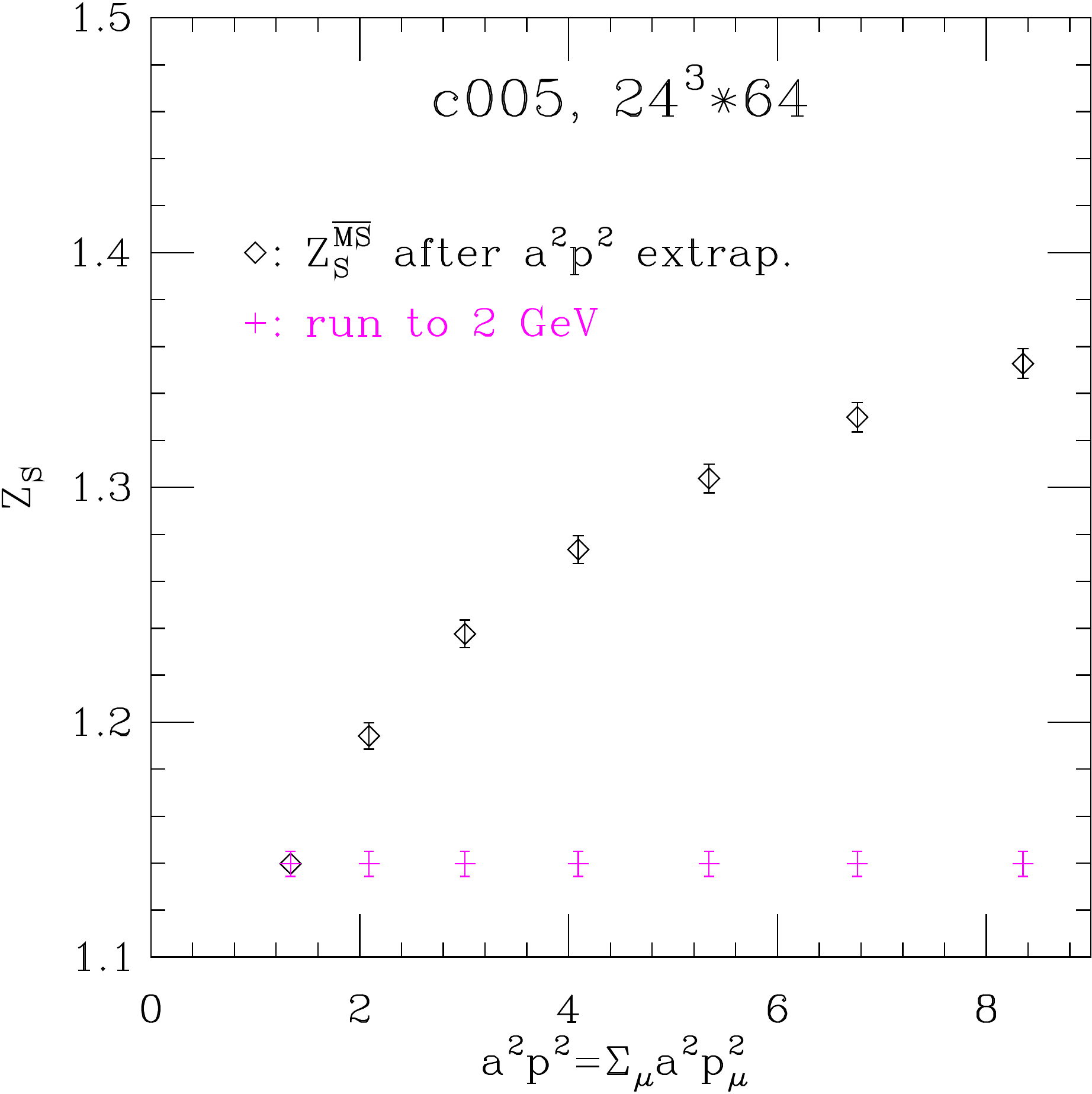}
\end{center}
\caption{Self-consistency check for the $a^2p^2$ extrapolation in the $\msbar$ scheme. See text.}
\label{fig:zs_check}
\end{figure}
We do a self-consistency check in Fig.~\ref{fig:zs_check} for the $a^2p^2$ extrapolation after the running in the
$\msbar$ scheme. The black diamonds in the graph are
$Z_S^{\msbar}(p)$ at $p=2, 2.5, 3, 3.5, 4, 4.5, 5$ GeV obtained from $a^2p^2$ extrapolations after the running in the $\msbar$ scheme
as described above for getting $Z_S^\msbar (p=2$ GeV$)$.
If the extrapolation works in reducing discretization errors, then the black diamonds should be well described by perturbative running in the
$\msbar$ scheme.
We run down the black diamonds to $2$ GeV using the 4-loop perturbative running in the $\msbar$ scheme. The results are the magenta pluses which
lie on a horizontal line within errors. This indicates that the $a^2p^2$ extrapolation can indeed reduce $\mathcal{O}(a^2p^2)$ discretization effects
and the higher order effects are small.

\begin{figure}
\begin{center}
\includegraphics[width=0.4\textwidth]{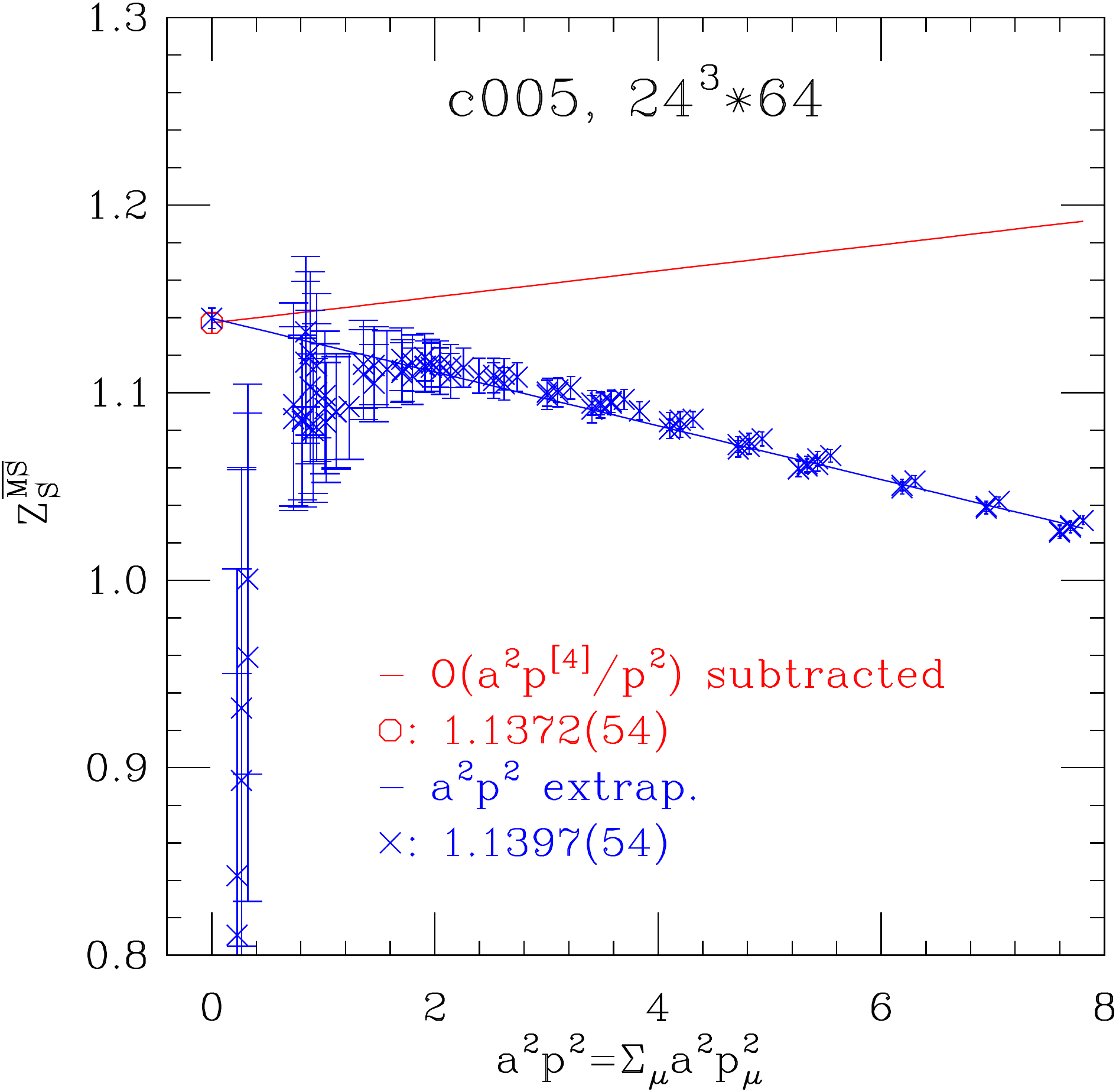}
\end{center}
\caption{Comparison of the three term fit Eq.(\ref{eq:a2p4p2}) with the $a^2p^2$ extrapolation.
The blue crosses are the results evolved to 2 GeV in the $\msbar$ scheme as a function of the initial renormalization scale.
The red line shows the three term fit function with the third term $c_2\frac{a^2p^{[4]}}{p^2}$ subtracted. The blue line is the
$a^2p^2$ extrapolation.}
\label{fig:zs_a2p2_a2p4p2}
\end{figure}

The blue crosses in Fig.~\ref{fig:zs_msbar}(left panel) do not necessary have a same $p^{[4]}/(p^2)^2$ value because the momentum modes we use
are not exactly in a same direction.
To see how the difference in $p^{[4]}/(p^2)^2$ affects our final result, we use the three term function
\begin{equation}
\mathcal{Z_S}=Z_S+c_1(a^2p^2)+c_2\frac{a^2p^{[4]}}{p^2}
\label{eq:a2p4p2}
\end{equation}
to fit the blue crosses in Fig.~\ref{fig:zs_msbar}(left panel) with $a^2p^2>5$. Here other possible terms proportional to 
$a^2p^{[6]}/(p^2)^2$, $a^4(p^2)^2$, etc. are ignored since Eq.(\ref{eq:a2p4p2}) can already fit the data.
The comparison of the three term fit with the $a^2p^2$ extrapolation
is shown in Fig.~\ref{fig:zs_a2p2_a2p4p2}.
Compared with the simple $a^2p^2$ extrapolation, the three term fit decreases $\chi^2/\mbox{dof}$ visibly.
The red line in Fig.~\ref{fig:zs_a2p2_a2p4p2} shows the fit function with the third term $c_2\frac{a^2p^{[4]}}{p^2}$ subtracted.
The $\mathcal{O}(a^2p^{[4]}/p^2)$ effects are not small since the red line is quite different from the blue data points.
However from the three term fit we get $Z_S^\msbar(2\mbox{ GeV})=1.1372(54)$, which is in good agreement with $1.1397(54)$ 
from the $a^2p^2$ extrapolation. This means with our statistical errors and with the condition in Eq.(\ref{eq:p4p22}), the effects due to the difference 
in the directions of the momenta can be ignored.

Comparing the slope in our $a^2p^2$ extrapolation with that in figure 2 of Aoki~\cite{Aoki:2010yq} (with NNNLO perturbative running),
we find a larger $a^2$ effect in our data. Similar size of slopes were also seen in Refs.~\cite{Aubin:2009jh,Gupta:2014dla},
where gauge fields were also smeared, for renormalization constants. It is possible that our gauge smearing is related to the size of the slope
in the $a^2p^2$ extrapolation. It is discussed in Ref.~\cite{Arthur:2013bqa} that link smearing may lower the upper end of the
RI-MOM window and enhance $a^2$ effects. A study to compare our results with thin link results would be interesting to understand better the slope.

The values of $Z_S^\msbar (2$ GeV$)$ on all ensembles are collected in Tab.~\ref{tab:zs}, where we have used
$a^2p^2>5$ for the $a^2p^2$ extrapolations on the $L=24$ lattices and $a^2p^2>3$ on the $L=32$ lattices.
\begin{table}
\begin{center}
\caption{$Z_S^\msbar(2$ GeV) on the $24^3\times64$ and $32^3\times64$ lattices.}
\begin{tabular}{ccccc}
\hline\hline
ensemble &   c02 & c01 & c005 & $m_{l}+m_{res}=0$  \\
$Z_S^\msbar(2$ GeV) &  1.1545(74) & 1.1361(82) & 1.1397(54) & 1.1272(87) \\
\hline
ensemble &   f008 & f006 & f004 & $m_{l}+m_{res}=0$ \\
$Z_S^\msbar(2$ GeV) &   1.074(10) & 1.0714(64) & 1.0574(65) & 1.0563(64) \\
\hline\hline
\end{tabular}
\label{tab:zs}
\end{center}
\end{table}

From the values on all six ensembles with different sea quark masses on the $L=24$ and $32$ lattices, we do a simultaneous linear extrapolation in the 
renormalized light sea quark mass 
to obtain $Z_S^\msbar$ in the sea quark massless limit. The fit function is
\begin{equation}
Z(m_l^R)=Z(0)+c\cdot m_l^R,\quad\mbox{where }m_l^R=(m_l+m_{res})Z_m^{sea}.
\label{eq:extrap_msea}
\end{equation}
Here $Z_m^{sea}=1.578(2)$ on the $L=24$ lattice and $1.573(2)$ on the $L=32$ lattice were given in Ref.~\cite{Aoki:2010dy}. The slopes of the two lines for
the coarse and fine lattices are required to be the same.

\begin{figure}
\begin{center}
\includegraphics[height=2.3in,width=0.4\textwidth]{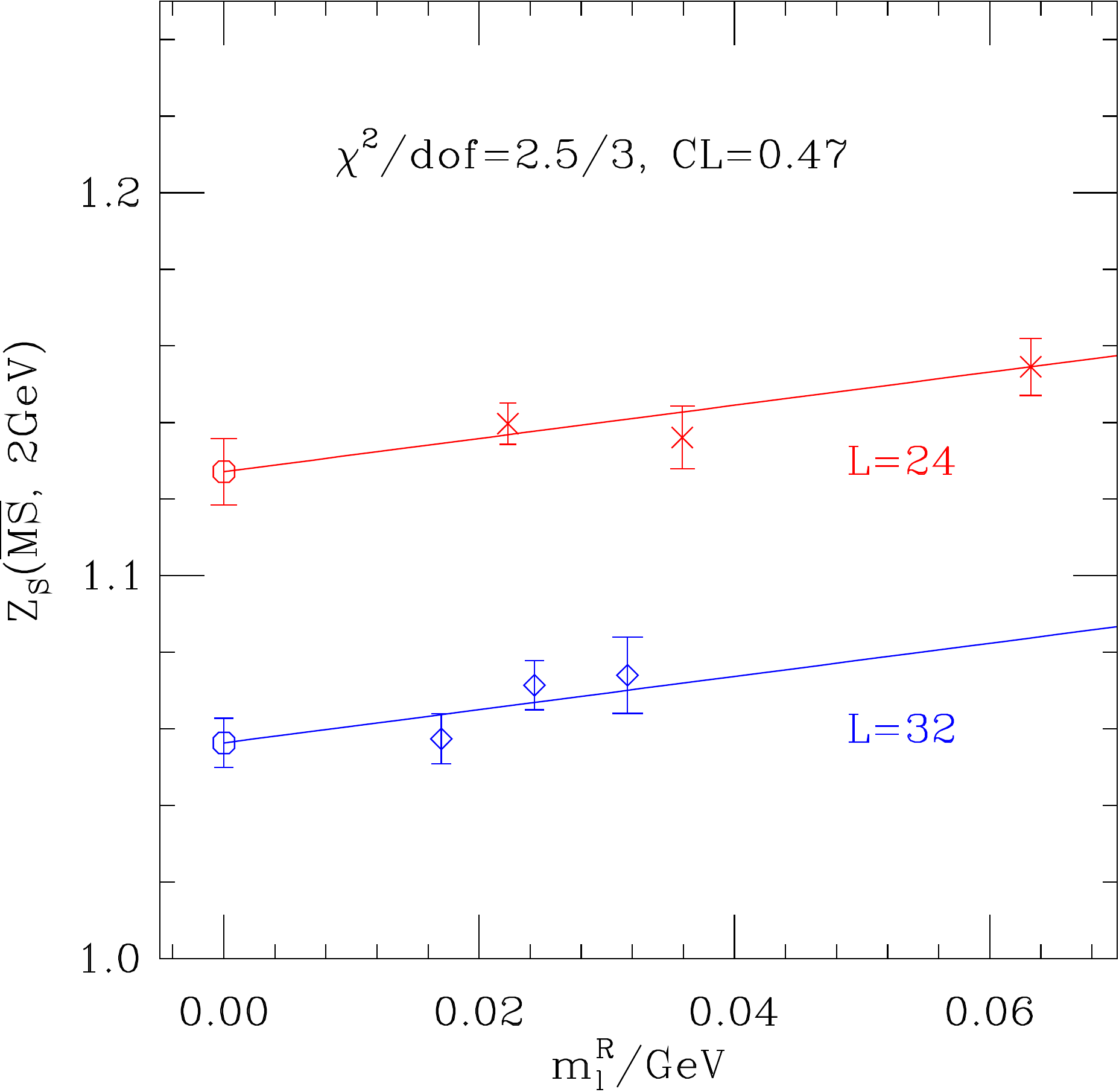}
\end{center}
\caption{Linear extrapolation of $Z_S^{\msbar}$ to the light sea quark massless limit.}
\label{fig:zs_extra_sea}
\end{figure}

The extrapolation is shown in Fig.~\ref{fig:zs_extra_sea}, which has a good $\chi^2/$dof. 
We do the simultaneous fit because the three light sea quark masses on the $L=32$ lattice are close to each other
and thus the data have less control on the slope.
Finally we get $Z_S^\msbar(L=24)=1.1272(87)$ and $Z_m^\msbar(L=24)=1/Z_S^{\msbar}=0.887(7)$ at 2 GeV. For the fine lattice we find
$Z_S^{\msbar}(L=32)=1.0563(64)$ and $Z_m^\msbar(L=32)=0.947(6)$.

We also did separate linear extrapolations in light sea quark masses on the coarse and fine lattices. 
The results are in agreement with those from the simultaneous fit. The change in the center values will be taken as one source of the
systematic errors as discussed below.

Besides the statistical error, we consider the following systematic errors of $Z_S$. The error budget of $Z_S$ in the chiral limit
is given in Tab.~\ref{tab:zs_error}.
\begin{table}
\begin{center}
\caption{Error budget of $Z_S^\msbar(2$ GeV) in the chiral limit}
\begin{tabular}{lcc}
\hline\hline
Source & Error (\%,L=24) & Error (\%,L=32)  \\
\hline
Statistical & 0.8 & 0.6 \\
\hline
Truncation (RI to $\msbar$)  &  1.5 & 1.4 \\
Coupling constant  & 0.3 & 0.3 \\
Perturbative running &  $<$0.02 & $<$0.02 \\
Lattice spacing & 0.5 & 0.4 \\
Fit range of $a^2p^2$ & 0.4 & 0.1 \\
Extrapolation in $m_l^R$ & 0.2 & 1.8 \\
Total systematic uncertainty & 1.7 & 2.3 \\
\hline\hline
\end{tabular}
\label{tab:zs_error}
\end{center}
\end{table}

First of all, high order terms that were ignored in the conversion ratio, Eq.(\ref{eq:ratio_zs_zp}), from the 
RI scheme to the $\msbar$ scheme give truncation errors. To reduce this error, one uses $Z_S^{RI}$ at large $a^2p^2$. 
In our work, we use $a^2p^2>5$ on the $L=24$ lattice
which means $p>3.87$ GeV. On the $L=32$ lattice, we use $a^2p^2>3$ or $p>4.02$ GeV.
At $p=4$ GeV, the numerical value of Eq.(\ref{eq:ratio_zs_zp}) is
\begin{eqnarray}
\frac{Z_S^{\rm RI}}{Z_S^{\overline{\rm MS}}}(p=4\mbox{ GeV},n_f=3)&=&1-0.424\alpha_s-0.827\alpha_s^2-1.944\alpha_s^3+\cdots\nonumber\\
&=&1-0.092-0.039-0.020+\cdots,
\end{eqnarray}
where we have used $\alpha_s^\msbar(4\mbox{ GeV})=0.2160$.
The $\mathcal{O}(\alpha_s^3)$ term is about 2.4\% of the total ratio.
The ignored $\mathcal{O}(\alpha_s^4)$ term is further suppressed by a factor of $\alpha_s$.
Assuming its coefficient is 3 times large as that for the $\mathcal{O}(\alpha_s^3)$ term, we get a $\sim1.5\%$ truncation error.

The uncertainty of the coupling constant $\alpha_s$ in Eq.(\ref{eq:ratio_zs_zp}) is another source of error. 
If we use $\Lambda_{QCD}^\msbar=349$ MeV instead of
339 MeV to evaluate $\alpha_s$, the center value of $Z_S^\msbar(2$ GeV$)$ changes by 0.3\% on both lattices.

The perturbative running of $Z_S^\msbar$ from an initial scale $p$ to 2 GeV uses 4-loop results of the anomalous dimension. The $\mathcal{O}(\alpha_s^4)$ term contributes
less than 0.02\% to the total running in our range of the initial scale $a^2p^2$. Thus this systematic error can be safely ignored.

To determine where 2 GeV is, we need the values of our lattice spacings. The variation of lattice spacings in the range of one sigma leads to $\sim0.5$\% change in
$Z_S^\msbar(2$ GeV$)$.

In the extrapolation of $Z_S^\msbar(2$ GeV;$ a^2p^2)$ to reduce $\mathcal{O}(a^2p^2)$ discretization errors, the fit range of $a^2p^2$ introduces 0.4\% error on
the $L=24$ lattice and 0.1\% error on the $L=32$ lattice. Here we vary $a^2p^2>5$ to $>4$ on the $L=24$ lattice and $a^2p^2>3$ to $>2$ on the $L=32$ lattice.

Finally, we consider the error due to the extrapolation in the light sea quark mass. As mentioned above, one can do separate and simultaneous fits to the data on the coarse and fine lattices.
The difference in the center values is taken as a systematic error.

In total, adding all systematic errors quadratically we find 1.7\% error for $Z_S$ on the coarse lattice and 2.3\% on the fine lattice. 
Putting the statistical
and systematic errors together, we have $Z_S^\msbar(2$ GeV)=1.127(9)(19) on the coarse lattice and 1.056(6)(24) on the fine lattice. The statistical error is
much smaller than the systematic error.

\subsection{Step scaling function of the quark mass}
\label{sec:step}
We can use the above obtained $Z_S^\msbar(2$ GeV) to determine strange and charm quark masses~\cite{Yang:2014taa} in the $\msbar$ scheme. 
Another way is to first
consider the continuum limit of renormalized RI data (quark mass, for example) at a fixed physical scale and then convert to the $\msbar$ scheme
by perturbation theory at a high enough scale. This strategy was used in, for example, Ref.~\cite{Aoki:2010pe}. In this way, $a^2p^2$ extrapolation
of the renormalization constants at large $p$ is not used to avoid possible lattice artifacts: the upper edge of the RI-MOM window may be
reduced by link smearing~\cite{Arthur:2013bqa}.

To use the above strategy to determine quark masses, we need the RI-MOM step scaling function in the continuum limit to run up to a high scale where
perturbative conversion ratio to the $\msbar$ scheme can be used.
Following Refs.~\cite{Arthur:2010ht,Arthur:2013bqa}, we calculate the step scaling function in the RI-MOM scheme for the quark mass as below.
Define a ratio
\begin{equation}
R_{\mathcal{O}}(\mu,a,m_q)=\frac{\Gamma_A(\mu,a,m_q)}{\Gamma_{\mathcal{O}}(\mu,a,m_q)}=\frac{Z_{\mathcal{O}}(\mu,a,m_q)}{Z_A}.
\end{equation}
With $Z_A$ determined, for example, as in Sec.~\ref{sec:za_wi}, one can get the renormalization constant
\begin{equation}
Z_{\mathcal{O}}(\mu,a)=Z_A\lim_{m_q\rightarrow0}R_{\mathcal{O}}(\mu,a,m_q).
\end{equation}
A ratio of the $R_{\mathcal{O}}$'s at different scales is the step scaling function
\begin{equation}
\Sigma_{\mathcal{O}}(\mu,s\mu,a)=\lim_{m_q\rightarrow0}\frac{R_{\mathcal{O}}(s\mu,a,m_q)}{R_{\mathcal{O}}(\mu,a,m_q)}
=\lim_{m_q\rightarrow0}\frac{Z_\mathcal{O}(s\mu,a,m_q)}{Z_\mathcal{O}(\mu,a,m_q)}.
\end{equation}
Its continuum limit is 
\begin{equation}
\sigma_{\mathcal{O}}(\mu,s\mu)=\lim_{a\rightarrow0}\Sigma_{\mathcal{O}}(\mu,s\mu,a)=\frac{Z_\mathcal{O}(s\mu)}{Z_\mathcal{O}(\mu)}.
\end{equation}
For the quark mass renormalization, using $Z_m=1/Z_S$ we have
\begin{equation}
\Sigma_{m}(\mu,s\mu,a)=\lim_{m_q\rightarrow0}\frac{Z_{S}(\mu,a,m_q)}{Z_{S}(s\mu,a,m_q)}
=\frac{\lim_{m_q\rightarrow0}Z_{S}(\mu,a,m_q)}{\lim_{m_q\rightarrow0}Z_{S}(s\mu,a,m_q)}.
\label{eq:Sigma_m}
\end{equation}
To calculate $\Sigma_{m}(\mu,s\mu,a)$ in the RI-MOM scheme, we use $Z_S^{\rm RI}$ which are already in the valance quark massless limit as
computed in Sec.~\ref{sec:zs_ri}, for example, the black diamonds in the left panel of Fig.~\ref{fig:zs_msbar} for ensemble c005. After a linear
extrapolation to the light sea quark massless limit ($m_l+m_{res}=0$) of those $Z_S^{\rm RI}$, 
we obtain $\Sigma_{m}(\mu,s\mu,a)$ by using interpolations explained below and Eq.(\ref{eq:Sigma_m}).

The scales $p$ in physical units for $Z_S^{\rm RI}(a^2p^2,a)$ at our two lattice spacings do not exactly match in the data. 
Therefore we interpolate the lattice data $Z_S^{\rm RI}(a^2p^2,a)$ in $a^2p^2$ with the ansatz
\begin{equation}
\frac{c_{-1}}{a^2p^2}+c_l\ln(a^2p^2)+c_0+c_1(a^2p^2).
\label{eq:zs_ri_ansatz}
\end{equation}
The first term in the above comes from the $1/p^2$ behavior of possible non-perturbative effects at low momenta. $c_1(a^2p^2)$ takes care of the discretization
effects. The other terms mimic the running of the operator. We fit our data with the above ansatz in the whole range of momenta available. Then we
interpolate to some physical scales $p=\mu$, which are chosen to be the same at the two lattice spacings.

\begin{figure}
\begin{center}
\includegraphics[width=0.4\textwidth]{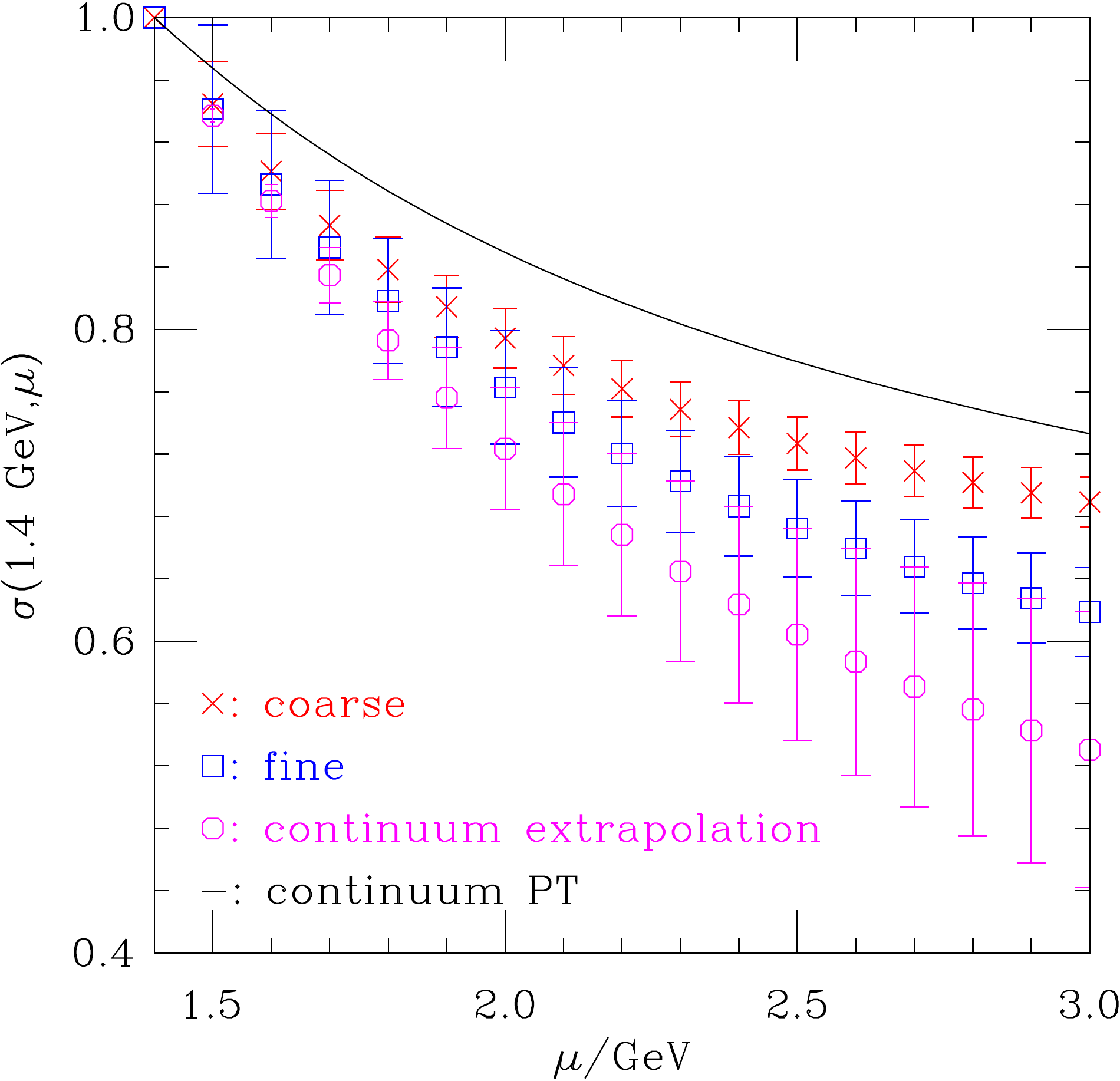}
\end{center}
\caption{The step scaling function for $Z_m$ and its value extrapolated to the continuum in the RI-MOM scheme.}
\label{fig:step_m}
\end{figure}

The step scaling function of the mass in the RI-MOM scheme from $\mu=1.4$ GeV to a higher scale $s\mu$ which is in the range [1.4 GeV, 3 GeV]
is plotted in Fig.~\ref{fig:step_m}.
We choose the relatively small value 1.4 GeV to follow Ref.~\cite{Arthur:2013bqa}. 
Another reason is that $\mathcal{O}(a^2\mu^2)$ discretization errors are smaller at lower $\mu$.
In the graph the red crosses are the step
scaling function on the coarse lattice, the blue squares are on the fine lattice. 
Then we consider the continuum limit of $\Sigma_{m}(\mu,s\mu,a)$ at given $\mu$ and $s\mu$.
The magenta octagons are from linear extrapolations in $a^2$ to the
continuum limit. Because there are only two lattice spacings, the extrapolated results have large error bars. The black curve is the 4-loop
perturbative result for the RI-MOM scheme for comparison. 

$Z_S^{\rm RI}(\mu)$ at the physical scale $\mu=1.4\mbox{ GeV}$ (no $a^2p^2$ extrapolation as in Sec.~\ref{sec:zs_ri} is performed) 
at the two lattice spacings can be obtained by interpolations.
Fitting our RI scheme data (for example, black diamonds in the left panel of Fig.~\ref{fig:zs_msbar}) with 
the ansatz Eq.(\ref{eq:zs_ri_ansatz}) in the whole range of momenta available, 
we get $Z_S^{\rm RI}(1.4$ GeV$)=0.7317(72)$ for ensemble c005. Similarly we obtain the values on other ensembles. Results on all ensembles
are given in Tab.~\ref{tab:zs_ri}.
\begin{table}
\begin{center}
\caption{$Z_S^{\rm RI}(\mu=1.4\mbox{ GeV})$ on the $24^3\times64$ and $32^3\times64$ lattices.}
\begin{tabular}{ccccc}
\hline\hline
ensemble &   c02 & c01 & c005 & $m_{l}+m_{res}=0$  \\
$Z_S^{\rm RI}(1.4$ GeV) & 0.852(19) &  0.782(14) &  0.7317(72) & 0.653(14) \\
\hline
ensemble &   f008 & f006 & f004 & $m_{l}+m_{res}=0$ \\
$Z_S^{\rm RI}(1.4$ GeV) &  0.772(18)  & 0.682(10) &  0.6606(74) & 0.606(11) \\
\hline\hline
\end{tabular}
\label{tab:zs_ri}
\end{center}
\end{table}
A simultaneous fit to $Z_S^{\rm RI}$ using Eq.(\ref{eq:extrap_msea}) to go to the light sea quark massless limit gives us the numbers in the
last column of Tab.~\ref{tab:zs_ri}.

As we have calculated, the step scaling function in the continuum limit for the mass in the RI-MOM scheme is
\begin{equation}
\sigma_m(1.4\mbox{ GeV},2\mbox{ GeV})=\frac{Z_m^{RI}(2\mbox{ GeV})}{Z_m^{RI}(1.4\mbox{ GeV})}=0.723(39).
\label{eq:step}
\end{equation}
This can be used to run up to 2 GeV from 1.4 GeV after one gets the RI-MOM scheme quark masses in the continuum limit.
The conversion ratio $Z_S^\msbar/Z_S^{RI}$ from the RI to the $\msbar$ scheme from Eq.(\ref{eq:ratio_zs_zp}) at 2 GeV is 1.289614, which
can then be used to obtain quark masses in the $\msbar$ scheme.

If we take the two numbers in the last column of Tab.~\ref{tab:zs_ri}, divide them by the number in Eq.(\ref{eq:step}) and convert to the
$\msbar$ scheme by using 1.289614, then we get 1.165(67) and 1.081(62) for the coarse and fine lattice respectively. They are in agreement with
the two numbers in the last column of Tab.~\ref{tab:zs} although here the error bar is large. 

\subsection{Pseudoscalar density}
\begin{figure}
\begin{center}
\includegraphics[height=2.3in,width=0.4\textwidth]{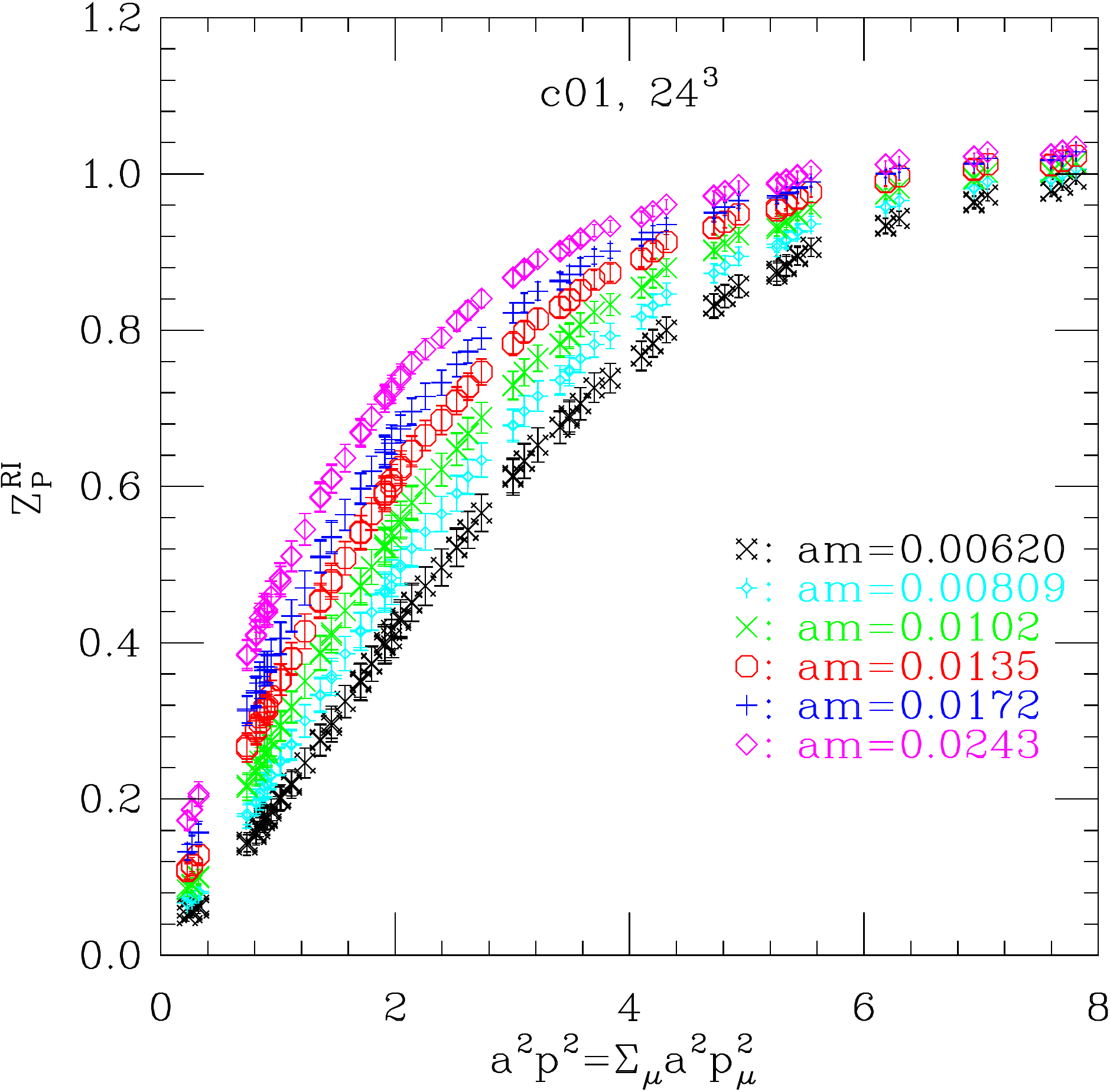}
\end{center}
\caption{An example of $Z_P^{RI}$ as a function of the momentum scale for ensemble c01 for different valence quark masses.}
\label{fig:zp_ri}
\end{figure}

The pseudoscalar renormalization constant $Z_P^{RI}$ from Eq.(\ref{eq:ri_condition}) is shown in Fig.~\ref{fig:zp_ri} for ensemble c01.
Because of the coupling to the Goldstone boson channel~\cite{Martinelli:1994ty}, the projected vertex function $\Gamma_P$ is divergent in
the valence quark massless limit. This non-perturbative contamination is suppressed at large scale as $1/p^2$. 
The singular behavior in $Z_P^{RI}$ at small $a^2p^2$
as shown in Fig.~\ref{fig:zp_ri}
is due to this contamination. To remove this non-perturbative effect, we fit $1/Z_P^{RI}$ at each given $a^2p^2$ to the ansatz~\cite{Becirevic:2004ny}
\begin{equation}
Z_P^{-1}=\frac{A}{am_q}+B+C(am_q),
\label{eq:zp_chiral_ansatz}
\end{equation}
where $A, B$ and $C$ are three fit parameters.
Then $Z_P^{sub}=B^{-1}$ is the value we take in the valence quark chiral limit.

\begin{figure}
\begin{center}
\includegraphics[height=2.6in,width=0.49\textwidth]{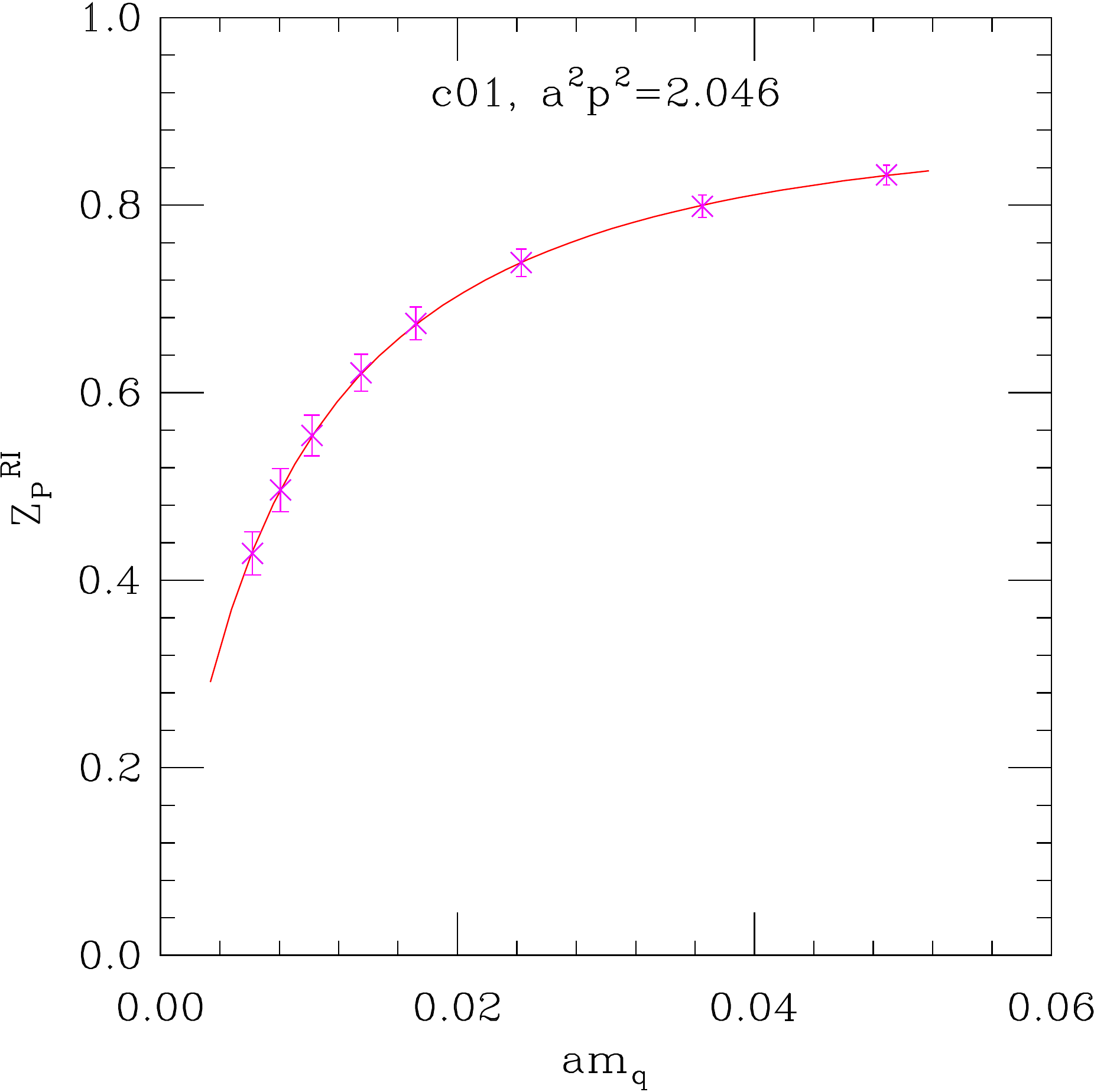}
\includegraphics[height=2.6in,width=0.49\textwidth]{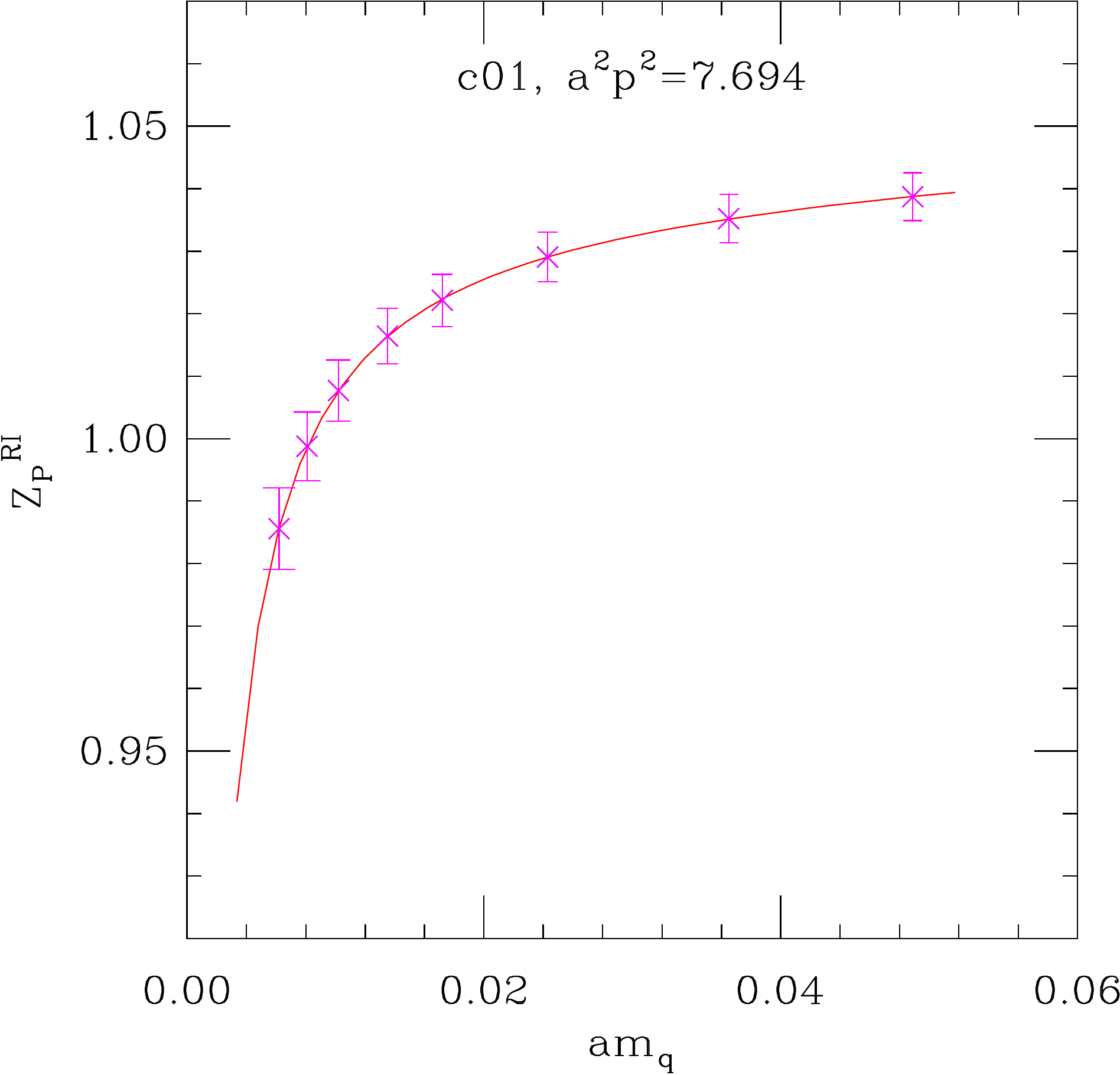}
\end{center}
\caption{Examples of fittings of $Z_P^{RI}$ to Eq.(\ref{eq:zp_chiral_ansatz}) for ensemble c01.}
\label{fig:zp_extrap_eg}
\end{figure}

\begin{figure}
\begin{center}
\includegraphics[height=2.3in,width=0.4\textwidth]{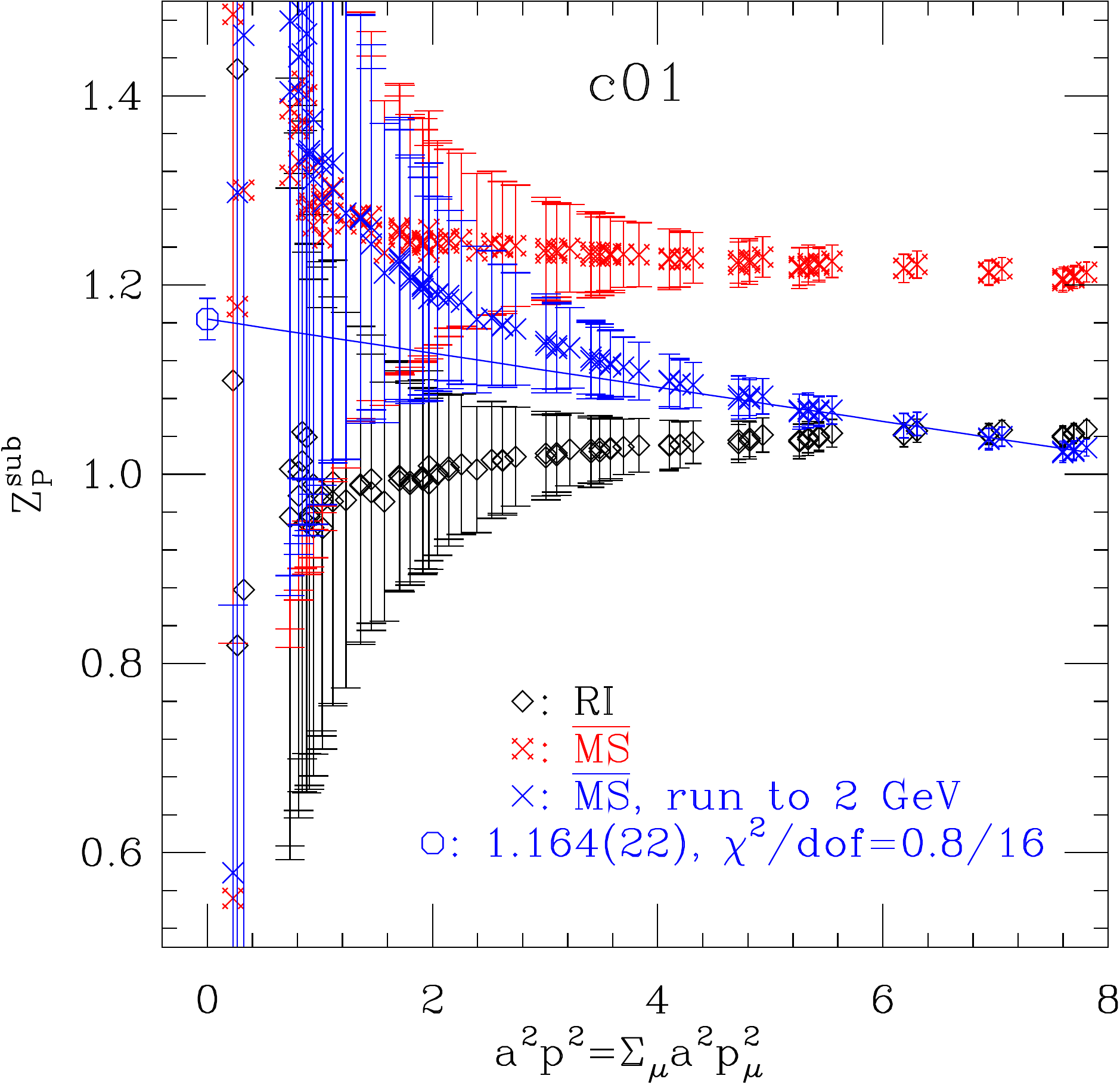}
\end{center}
\caption{The conversion and running of $Z_P^{sub}$ in the valence quark massless limit on ensemble c01. The black diamonds are the values in
the RI scheme. The red fancy crosses are those in the $\msbar$ scheme. The blue crosses are the results evolved to 2 GeV in the
$\msbar$ scheme as a function of the initial renormalization scale.}
\label{fig:zp_msbar}
\end{figure}

In Fig.~\ref{fig:zp_extrap_eg} we show some examples of the fitting of $Z_P^{RI}$ to Eq.(\ref{eq:zp_chiral_ansatz}) 
at some given $a^2p^2$. All the fittings have small $\chi^2/$dof. 
After obtaining $Z_P^{sub}$ in the RI scheme, we use Eq.(\ref{eq:ratio_zs_zp}) to convert to the $\msbar$ scheme. The results are shown by the
red fancy crosses in Fig.~\ref{fig:zp_msbar}.
Similar to the analysis of $Z_S$, we use the quark mass anomalous dimension to 
evolve $Z_{P,\msbar}^{sub}(a^2p^2)$ to 2 GeV in the $\msbar$ scheme and obtain the
blue crosses in Fig.~\ref{fig:zp_msbar}.
Then a linear fit in $a^2p^2$ (the blue solid line in Fig.~\ref{fig:zp_msbar}) to the data at $a^2p^2>5$ is used to extrapolate
away $\mathcal{O}(a^2p^2)$ discretization errors. We finally find $Z_{P,\msbar}^{sub}=1.164(22)$ at 2 GeV on ensemble c01.

\begin{figure}
\begin{center}
\includegraphics[height=2.3in,width=0.4\textwidth]{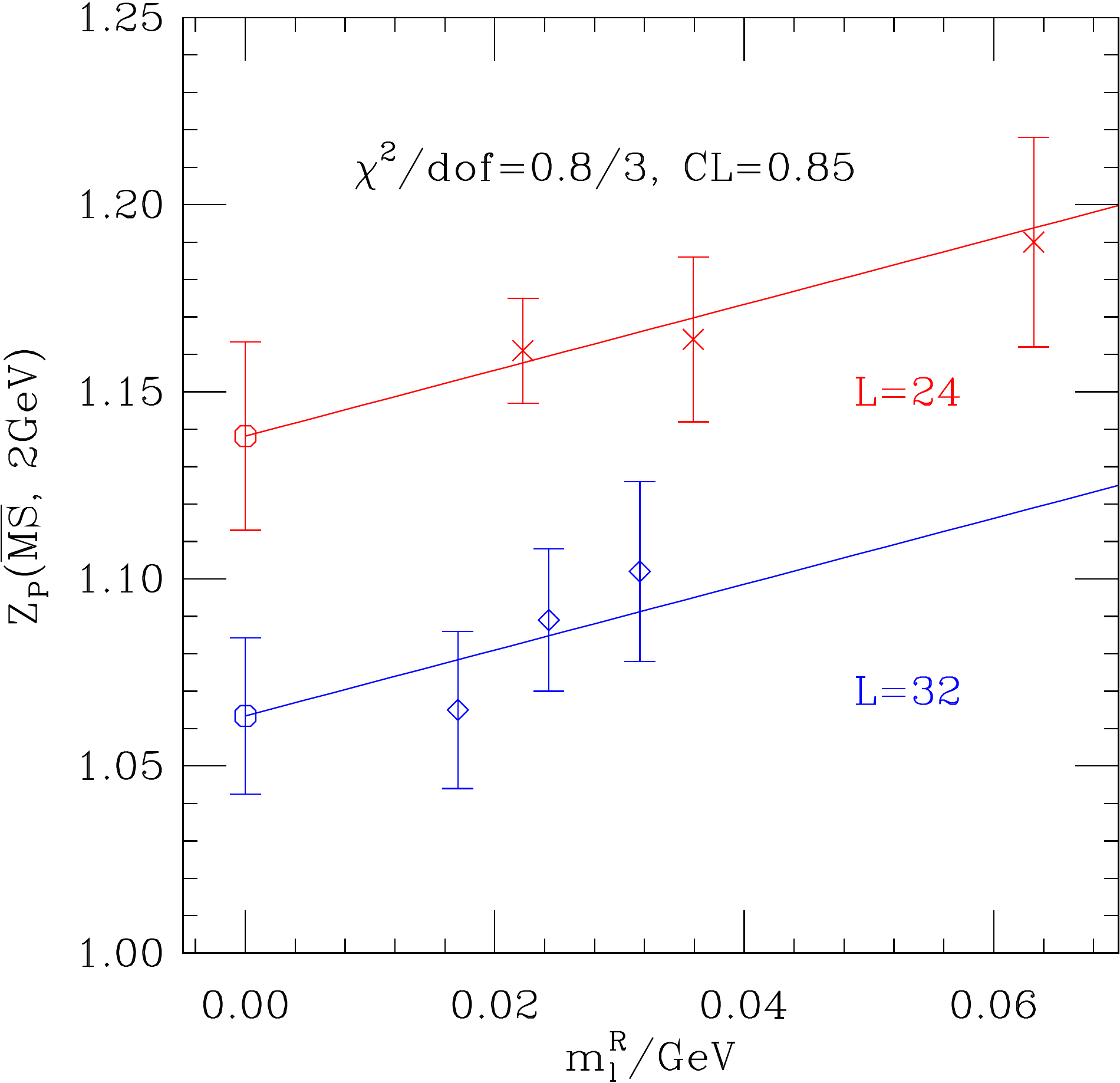}
\end{center}
\caption{Linear extrapolation of $Z_{P,\msbar}^{sub}$ to the light sea quark massless limit.}
\label{fig:zp_extra_sea}
\end{figure}

The values of $Z_{P,\msbar}^{sub}(2$ GeV$)$ on all ensembles are collected in Tab.~\ref{tab:zp}. 
\begin{table}
\begin{center}
\caption{$Z_{P,\msbar}^{sub}(2$ GeV) on the $24^3\times64$ and $32^3\times64$ lattices.}
\begin{tabular}{ccccc}
\hline\hline
ensemble &  c02 & c01 & c005 & $m_{l}+m_{res}=0$ \\
$Z_{P,\msbar}^{sub}(2$ GeV) &  1.190(28) & 1.164(22) & 1.161(14) & 1.138(25) \\
\hline
ensemble &  f008 & f006 & f004 & $m_{l}+m_{res}=0$ \\
$Z_{P,\msbar}^{sub}(2$ GeV) &  1.102(24) & 1.089(19) & 1.065(21) & 1.063(21) \\
\hline\hline
\end{tabular}
\label{tab:zp}
\end{center}
\end{table}
In the last column of Tab.~\ref{tab:zp}, the sea quark massless limit values of $Z_{P,\msbar}^{sub}$ are given. They are obtained from a
simultaneous linear extrapolation in the renormalized light sea quark mass to $Z_{P,\msbar}^{sub}$ on both $L=24$ and $32$ lattices. The extrapolation
is shown in Fig.~\ref{fig:zp_extra_sea} with the fit function given in Eq.(\ref{eq:extrap_msea}).
Comparing the numbers in Tab.~\ref{tab:zp} with those in Tab.~\ref{tab:zs}, we see that $Z_S=Z_P^{sub}$ is well satisfied within errors.

Similar to the analysis for $Z_S$, we summarize the systematic errors as well as the statistical error in Tab.~\ref{tab:zp_error}.
\begin{table}
\begin{center}
\caption{Error budget of $Z_{P,\msbar}^{sub}(2$ GeV) in the chiral limit}
\begin{tabular}{lcc}
\hline\hline
Source & Error (\%,L=24) & Error (\%,L=32)  \\
\hline
Statistical & 2.2 & 2.0 \\
\hline
Truncation (RI to $\msbar$)  &  1.5 & 1.4 \\
Coupling constant  & 0.3 & 0.3 \\
Perturbative running &  $<$0.02 & $<$0.02 \\
Lattice spacing & 0.5 & 0.4 \\
Fit range of $a^2p^2$ & 0.1 & 0.1 \\
Extrapolation in $m_l^R$ & 0.6 & 3.8 \\
Total systematic uncertainty & 1.7 & 4.1 \\
\hline\hline
\end{tabular}
\label{tab:zp_error}
\end{center}
\end{table}
Unlike $Z_S^\msbar$, the statistical error of $Z_{P,\msbar}^{sub}$ is about the same size as the systematic error.

\subsection{Vector current}
\begin{figure}
\begin{center}
\includegraphics[height=2.3in,width=0.4\textwidth]{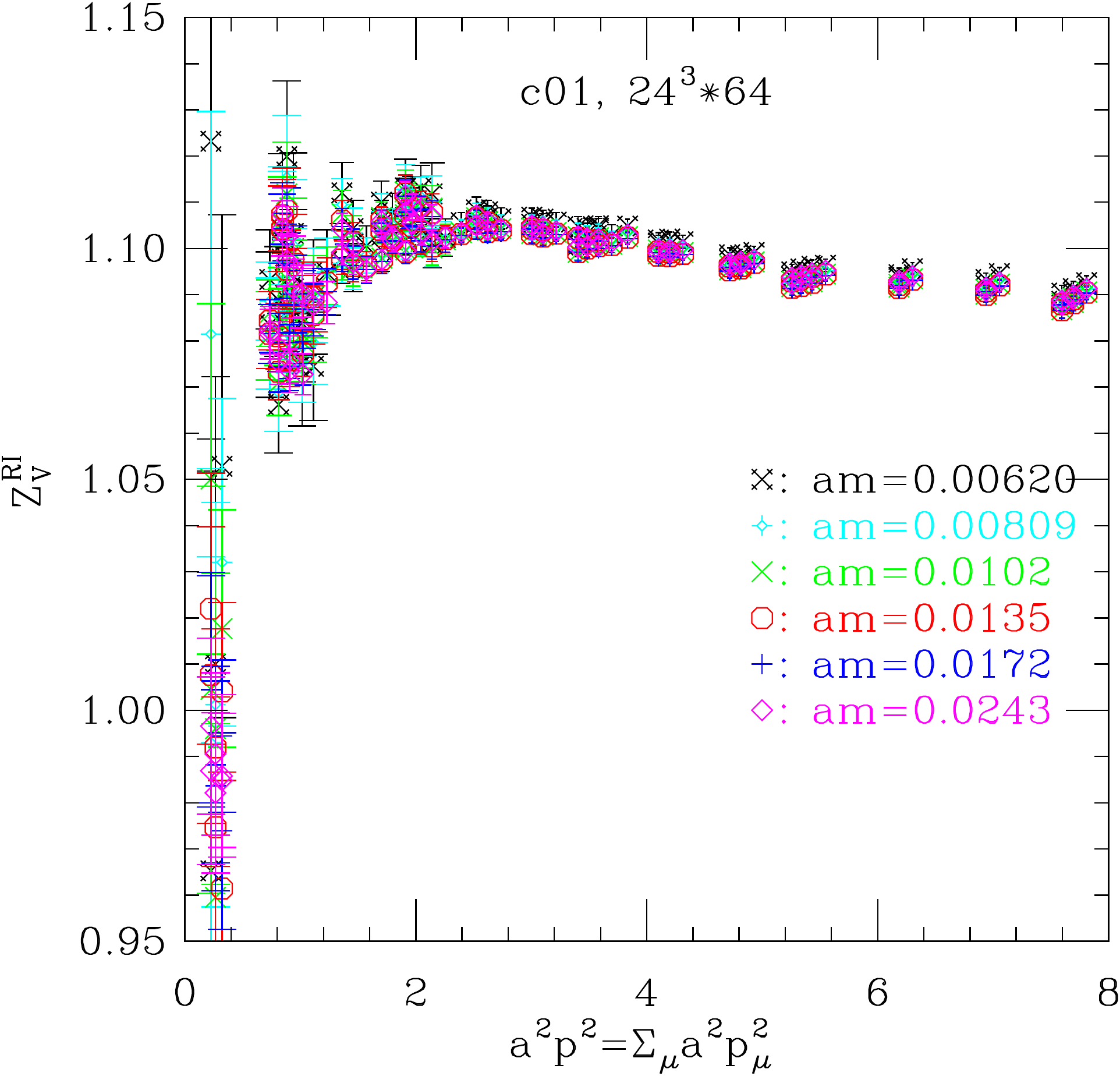}
\end{center}
\caption{Examples of $Z_V^{RI}$ as functions of the momentum scale for ensemble c01.}
\label{fig:zv_ri}
\end{figure}

The renormalization constant in the RI scheme for the local vector current for different valence quark masses on data ensemble c01
are shown in Fig.~\ref{fig:zv_ri}. Here in using Eq.(\ref{eq:ri_condition}), we have averaged $\mu=1,2,3,4$ for the vector current.
The valence quark mass dependence for $Z_V^{RI}$ is small so that the symbols in Fig.~\ref{fig:zv_ri} for different masses are almost
on top of each other. $Z_V^{RI}$ is scale independent when the renormalization scale is big. This is confirmed in
Fig.~\ref{fig:zv_ri}. At scales $a^2p^2>\sim3$, $Z_V^{RI}$ is flat up to discretization errors.

\begin{figure}
\begin{center}
\includegraphics[height=2.6in,width=0.49\textwidth]{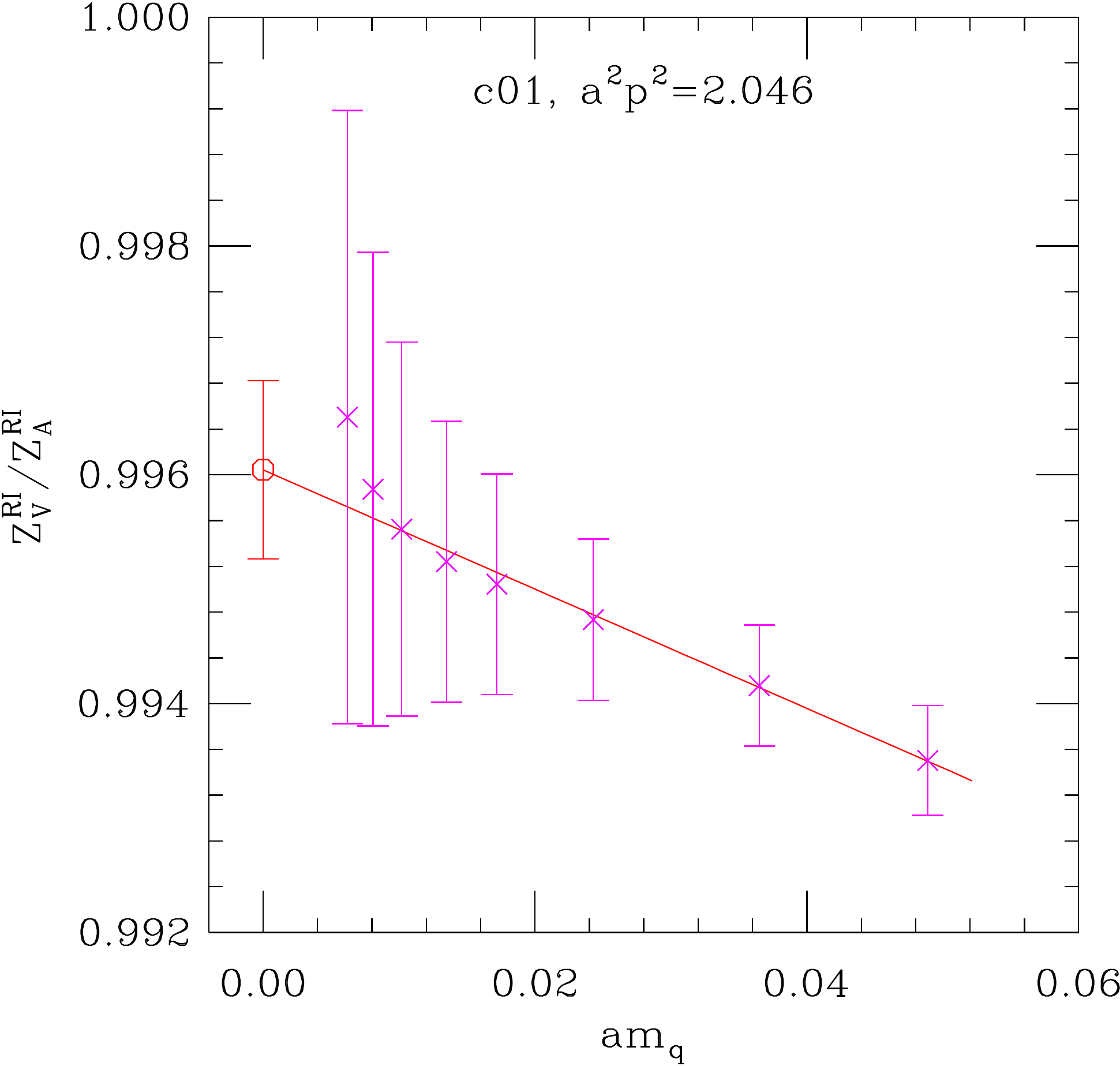}
\includegraphics[height=2.6in,width=0.49\textwidth]{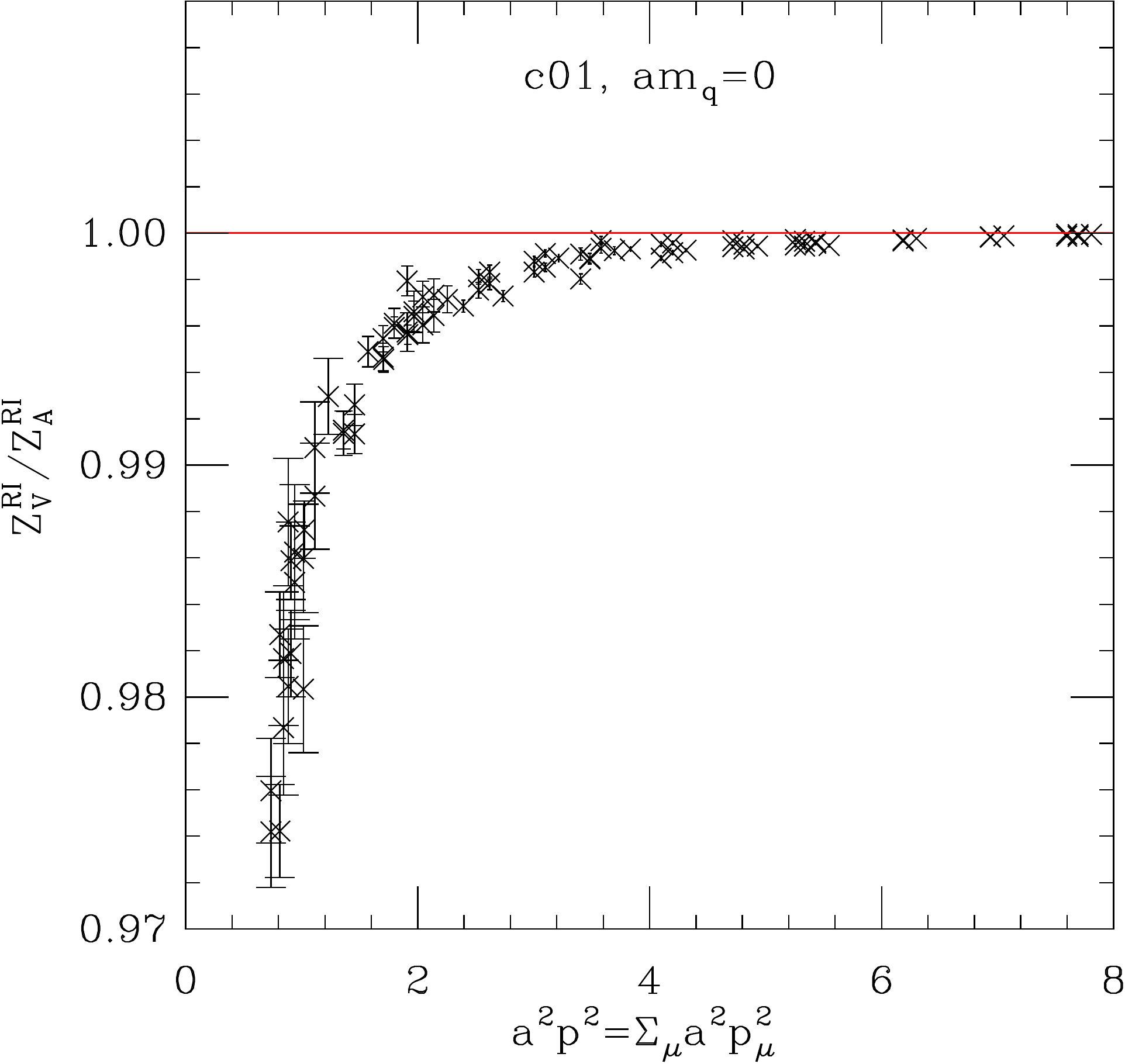}
\end{center}
\caption{Left panel: $Z_V^{RI}/Z_A^{RI}$ against the valence quark mass from ensemble c01 at a certain momentum scale and a linear extrapolation to $am_q=0$. 
Right panel: $Z_V^{RI}/Z_A^{RI}$ in the valence quark massless limit as a function of the momentum scale for ensembles c01.}
\label{fig:zv_over_za_ri}
\end{figure}

In Fig.~\ref{fig:zv_over_za_ri}, the ratio $Z_V^{RI}/Z_A^{RI}$ for ensemble c01 is shown.
To go to the chiral limit, we use a linear extrapolation in valence quark mass for $Z_V^{RI}/Z_A^{RI}$.
The left panel in Fig.~\ref{fig:zv_over_za_ri} shows an example of such extrapolations.
As we can see on the right panel of Fig.~\ref{fig:zv_over_za_ri}, at large momentum scales, $Z_V^{RI}/Z_A^{RI}=1$, i.e., $Z_V^{RI}=Z_A^{RI}$ is satisfied as expected.

The results of $Z_V^{RI}/Z_A^{RI}$ for other five ensembles are similar to those for ensemble c01.

\section{Summary}\label{sec:summary}
In this work, we obtain the renormalization constants for quark bilinear operators for the setup of overlap valence quark on 2+1-flavor domain wall fermion
configurations. We calculate those constants non-perturbatively by using Ward identity and the RI-MOM scheme. The matching factors from the lattice to the continuum $\msbar$ 
scheme for the scalar, pseudoscalar, vector and axial vector currents are obtained. 
$Z_S=Z_P$ and $Z_V=Z_A$ are confirmed for overlap fermions. 
The step scaling function of quark masses in the RI-MOM scheme is also calculated. By using the step scaling function in the continuum limit, 
the renormalized quark mass in the RI-MOM scheme can be run up to a high scale and then be converted to the $\msbar$ scheme.
Our main results are collected in Tabs.~\ref{tab:Za_WI_24_32},~\ref{tab:zs},~\ref{tab:zs_error},~\ref{tab:zs_ri},~\ref{tab:zp} and~\ref{tab:zp_error}. These matching factors
are important components in lattice determination of physical quantities such as quark masses, quark condensate and pseudoscalar meson decay constants.

The statistical error of $Z_S$ can reach less than one percent, which is much smaller than its systematic error.
A big contribution of the systematic error comes from the perturbative conversion ratio from the RI-MOM scheme to the $\msbar$ scheme. 
The RI-SMOM scheme~\cite{Sturm:2009kb} was shown
to have conversion ratios which converge much faster~\cite{Gorbahn:2010bf,Almeida:2010ns} and smaller non-perturbative effects. 
In the RI-SMOM scheme, the momentum magnitudes of the Green functions of the relevant operators are symmetric.
However in this work our boundary condition in the time direction is anti-periodic. This limits the number of symmetric momentum combinations (actually we 
cannot have exact symmetric momentum combinations). To shrink the systematic error, one can use a periodic boundary condition in the time direction or
twisted boundary conditions~\cite{Arthur:2010ht} with the RI-SMOM scheme.

\section*{Acknowledgements}
We thank RBC-UKQCD collaboration for sharing the domain wall fermion configurations. ZL thanks Thomas DeGrand, Ron Horgan and Olivier P\`ene for useful 
discussions. This work is partially supported by U.S.\ DOE Grant No.\ DE-FG05-84ER40154 and by the National Science Foundation of China (NSFC) under Grants 11105153,
10835002, 11075167, 11275169 and 11335001. ZL is partially supported by the Youth Innovation Promotion Association of CAS. 
YC and ZL acknowledge the support of NSFC and DFG (CRC110).

\end{document}